\documentclass[12pt]{article}
\usepackage{amssymb,amsmath,epsfig}
\allowdisplaybreaks

\begin{document}
\title{\bf Effects of Non-linear Electrodynamics on Thermodynamics of Charged Black Hole}
\author{M. Sharif \thanks{msharif.math@pu.edu.pk} and Amjad Khan\thanks{amjadcp112@gmail.com}\\
Department of Mathematics, University of the Punjab,\\
Quaid-e-Azam Campus, Lahore-54590, Pakistan.}

\date{}
\maketitle

\begin{abstract}
This paper investigates thermodynamics, quasi-normal modes, thermal
fluctuations and phase transitions of Reissner-Nordstr\"om black
hole with the effects of non-linear electrodynamics. We first
compute the expressions for Hawking temperature, entropy and heat
capacity of this black hole and then obtain a relation between
Davies's point and quasi-normal modes with non-linear
electrodynamics. We also observe the effects of logarithmic
corrections on uncorrected thermodynamic quantities such as entropy,
Hawking temperature, Helmholtz free energy, internal energy, Gibbs
free energy, enthalpy and heat capacity. It is found that presence
of non-linear electrodynamic parameter induces more instability in
black holes of large radii. Finally, we analyze the phase
transitions of Hawking temperature as well as heat capacity in terms
of entropy for different values of charge ($q$), horizon radius
($r_{+}$) and coupling parameter ($\alpha$). We obtain that Hawking
temperature changes its phase from positive to negative for
increasing values of $q$ and $r_{+}$ while it shows opposite trend
for higher values of $\alpha$. The heat capacity changes its phase
from negative to positive for large values of charge, horizon radius
and coupling parameter.
\end{abstract}

\section{Introduction}

In general relativity, one of the most interesting subjects is the
study of final outcomes of self-gravitating astronomical objects.
Black hole (BH) is the completely collapsed structure of massive
star which is defined as a thermodynamical object with infinite
gravity such that nothing not even electromagnetic radiations such
as light escape from it. There are some analogies between classical
and BH laws of thermodynamics such as temperature is related to
surface gravity of BH, energy has analogy with mass and entropy has
resemblance with area of event horizon. These analogies motivated
Bekenstein \cite{1} to find out a relation between area and entropy
of BH. In this regard, he found that entropy is proportional to area
of the event horizon of BH and Hawking's discovery of black body
radiations further confirmed the validity of this relation.
Consequently, it is impossible to obtain thermal equilibrium between
BH and thermal radiations such as area of the event horizon of BH
can never decrease. In this regard, the area-entropy relation
proposed by Bekenstein needs to be corrected, leading to the concept
of thermal fluctuations and holographic principle \cite{2}.

Quasi-normal modes (QNMs) are the perturbed solutions of linearized
dynamical equations of BH characterized by complex eigenvalues where
the real parts represent frequency oscillations and imaginary parts
express damping modes. The origin of perturbations of BH is the
pioneering work of Regge and Wheeler \cite{3} which was further
extended by Zerilli \cite{4}. Vishveshwara \cite{5} was the first to
calculate QNMs by the scattering of gravitational waves from
Schwarzschild BH. Leaver \cite{6} formulated analytic representation
of QNM wave-functions and also evaluated gravitational QNMs of both
non-rotating and rotating BHs. Jing and Pan \cite{7} studied the
relationship between QNMs and second order phase transitions of the
Reissner-Nordstr\"om (RN) BH. They found that both imaginary and
real parts of QNM frequencies become oscillatory functions of charge
and also contain thermodynamic information of BH.

Konoplya and Zhidenko \cite{8} studied different aspects related to
the perturbations of BHs, i.e., decoupling of variables in perturbed
equations, anti-de Sitter (AdS) evaluation of QNMs, late-time tails,
holographic superconductors and gravitational stability. Konoplya
and Stuchlik \cite{9} calculated analytical relation for
gravitational QNMs in the eikonal limit and studied the null
geodesics of asymptotically flat BH. Breton et al. \cite{10}
discussed QNMs as well as absorption cross-sections for the
Born-Infeld-dS BHs and used Wentzel-Kramers-Brillouin approximation
as well as null geodesics to evaluate QNMs of massless scalar
fields. Ovgun and Jusufi \cite{10a} discussed the QNMs, stability,
greybody factor and absorption cross section of $f(R)$ gravity
minimally coupled to a cloud of strings in 2+1 dimensions. Ovgun et
al. \cite{10b} formulated the QNMs for two types of BHs such as
4-dimensional dS and 5-dimensional Schwarzschild-AdS BHs by using
feedforward neural network method and found that obtained results
are similar to previous ones calculated from other methods.

Churilova \cite{11} found analytical QNMs of BHs in different
theories and added corrections due to deviations from Einstein
theory. He also derived a general relation for analytical evaluation
of the eikonal QNMs for asymptotically flat metrics with small
deviations from the Schwarzschild geometry. Sakalli et al.
\cite{11b} evaluated the QNMs and solution of Klein-Gordon equation
in Born-Infeld dilaton spacetime with cosmic string. Wei and Liu
\cite{11c} studied analytical relation between Davies point and QNMs
for RN BH in the eikonal limit and concluded that QNMs can be
obtained from null geodesics using the angular velocity as well as
Lyapunov exponent of photon sphere.

Fluctuations in compact objects due to statistical perturbations are
known as thermal fluctuations which has significant impact on the
geometry of BH. It is believed that Hawking radiations reduce the
size of BH which leads to increase its temperature. Faizal and
Khalil \cite{13} studied the effect of logarithmic corrections on
thermodynamics of three BHs, i.e., RN, Kerr as well as charged AdS
BHs and found that these BHs produce remnants in all cases.
Pourhassan and Faizal \cite{14} analyzed the effect of thermal
fluctuations on thermodynamic quantities of a small singly spinning
Kerr-AdS BH. They found that these fluctuations correct the entropy
of BH by a logarithmic correction term and also noted that
logarithmic corrections are very important for sufficiently small
BHs.

Jawad and Shahzad \cite{15} discussed the effects of thermal
fluctuations on non-minimal regular BHs with cosmological constant
and found that BHs are stable both locally as well as globally for
large values of cosmological constant. Zhang \cite{16} analyzed the
effect of first order correction terms to the entropy of RN-AdS as
well as Kerr-Newman-AdS BHs and concluded that corrected entropy
only affects the thermodynamic potentials of small BHs. Pradhan
\cite{16a} studied thermodynamics and thermal fluctuations for
charged accelerating BHs and found that such BHs are stable as well
as second order phase transitions occur in them.

Davies \cite{17} developed thermodynamical theory of BHs and found
that phase transition occurs in Kerr-Newman BHs. Hawking and Page
\cite{18} determined the presence of phase transition in
Schwarzschild-AdS BH. Biswas and Chakraborty \cite{19} investigated
that whether phase transition is possible in Horava Lifshitz gravity
under the consideration of classical and topological choices of BH
thermodynamics. They concluded that phase transition occurs in
Horava Lifshitz gravity from stable to unstable phase for increasing
values of radius. Kubiznak and Mann \cite{20} analyzed the first
order small-large BHs phase transition in charged AdS BH which is
analogous to liquid-gas phase transitions of fluids. Tharanath et
al. \cite{21} studied thermodynamical properties as well as phase
transition of regular BHs and found that regular BHs undergo second
order phase transition.

Chaturvedi et al. \cite{22} investigated phase transition in the
framework of thermodynamic geometry for charged AdS BH and found
first order phase transition for fixed electric charge. Wei et al.
\cite{23} studied thermodynamics and found first order phase
transition for charged AdS BHs in five-dimensions. Ovgun \cite{23a}
studied the thermodynamics as well as phase transition of a specific
charged AdS type BH in $f(R)$ gravity coupled with Yang-Mills field
by considering cosmological constant as thermal pressure. Wei and
Liu \cite{24} discussed a relation between phase transition and null
geodesics of charged AdS BH. Saleh et al. \cite{25} examined
thermodynamics as well as phase transition of Bardeen BH surrounded
by quintessence and found that presence of quintessence induces
phase transition in considered BH. Kuang et al. \cite{25a} evaluated
thermal quantities of AdS non-linear electrodynamic BH. Bhatacharya
et al. \cite{26} provided a general criteria to get information and
other various results with an extremal limit of BH spacetime. They
also evaluated critical values of second order phase transition
without considering any specific BH and showed that these values are
in agreement with those of any specific BH cases.

Gonzalez et al. \cite{29} studied thermodynamics and stability of
charged BHs with non-linear electrodynamics and found that small BHs
are stable locally. Balart and Vagenas \cite{30} derived the
solutions of many regular BHs with non-linear electrodynamics and
verified that some of them show asymptotic behavior to RN BH.
Dayyani \cite{30a} studied thermodynamical properties and phase
transition of dilaton BH with non-linear electrodynamics and
observed zeroth order phase transition. Yu and Gao \cite{31} derived
the exact solutions of RN BH with non-linear electrodynamics. Javed
et al. \cite{32} observed the effect of non-linear electrodynamics
on weak field deflection angle of RN BH. Recently, Fauzi and
Ramadhan \cite{32a} discussed thermodynamics of charged BHs with
non-linear electrodynamics and verified the first law of BH
thermodynamics.

In this paper, we study thermodynamical quantities, QNMs, thermal
fluctuations and phase transition for RN BH with non-linear
electrodynamic effects. The paper is outlined as follows. In section
\textbf{2}, we discuss thermodynamics as well as thermal stability
for the considered BH by using heat capacity. Section \textbf{3} is
devoted to examine the relation between QNMs and Davies point. In
section \textbf{4}, we study the effects of thermal fluctuations on
uncorrected thermodynamical quantities and section \textbf{5}
analyzes phase transitions in terms of entropy. In the last section,
we conclude all the results.

\section{Thermodynamics}
\begin{figure}\center
\epsfig{file=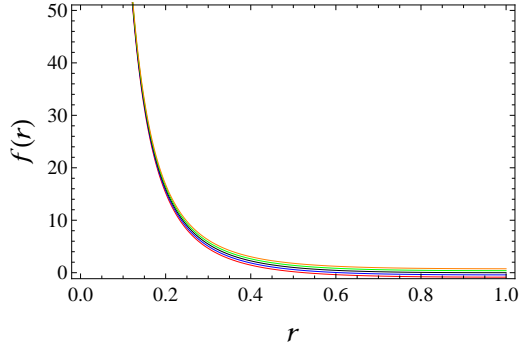,width=0.5\linewidth} \caption{Plot of the metric
function versus $r$ for $M=1=q$ with $\alpha=-0.4$(red), -0.2(blue),
0(black),  0.2(green) and 0.4(orange).}
\end{figure}

In this section, we discuss thermodynamics of RN BH with non-linear
electrodynamic effects such as Hawking temperature and heat
capacity. We also use derived expression of heat capacity to explore
the thermal stable configuration of the system. The line element of
considered BH is given as \cite{32}
\begin{equation}\label{1}
ds^{2}=f(r)dt^{2}-\frac{dr^{2}}{f(r)}-r^{2}d\theta^{2}-r^{2}\sin^{2}\theta
d\phi^{2},
\end{equation}
with
\begin{equation}\label{2}
f(r)=1-\frac{2 M}{r}+\frac{q^2}{r^2}+2 \alpha  q-\frac{\alpha ^2
r^2}{3},
\end{equation}
$M$, $q$ and $\alpha$ represent mass of BH, charge and coupling
constant, respectively. The line element (1) reduces to the RN BH
when $\alpha=0,~q\neq0$, Schwarzschild metric for $\alpha=0,~q=0$
and it becomes Schwarzschild BH with the effects of non-linear
electrodynamics for $\alpha\neq0,~q=0$. In Figure \textbf{1}, the
graphical analysis of metric function (\ref{2}) shows the appearance
of naked singularities for negative as well as positive values of
coupling parameter. Setting $f(r_{+})=0$, the mass of BH in terms of
$r_{+}$ can be obtained as
\begin{equation}\label{3}
M=\frac{3 q^2+6 \alpha  q r_{+}^2-\alpha ^2 r_{+}^4+3
r_{+}^2}{6r_{+}}.
\end{equation}
where $r_{+}$ is the horizon radius of BH. To obtain the expression
for Hawking temperature, we first evaluate surface gravity of the BH
\begin{equation}\nonumber
\kappa=-\frac{\left(q-\alpha  r_{+}^2+r_{+}\right) (q-r_{+} (\alpha
r_{+}+1))}{2 r_{+}^3}.
\end{equation}
Consequently, the Hawking temperature $(T=\frac{\kappa}{2\pi})$ is
\begin{equation}\label{3a}
T=-\frac{\left(q-\alpha  r_{+}^2+r_{+}\right) (q-r_{+} (\alpha
r_{+}+1))}{4\pi r_{+}^3}.
\end{equation}
The graphical sketch of Hawking temperature versus $\alpha$ is shown
in the left plot of Figure \textbf{2}. It is found that Hawking
temperature increases with the increasing values of coupling
parameter throughout the considered domain.
\begin{figure}\center
\epsfig{file=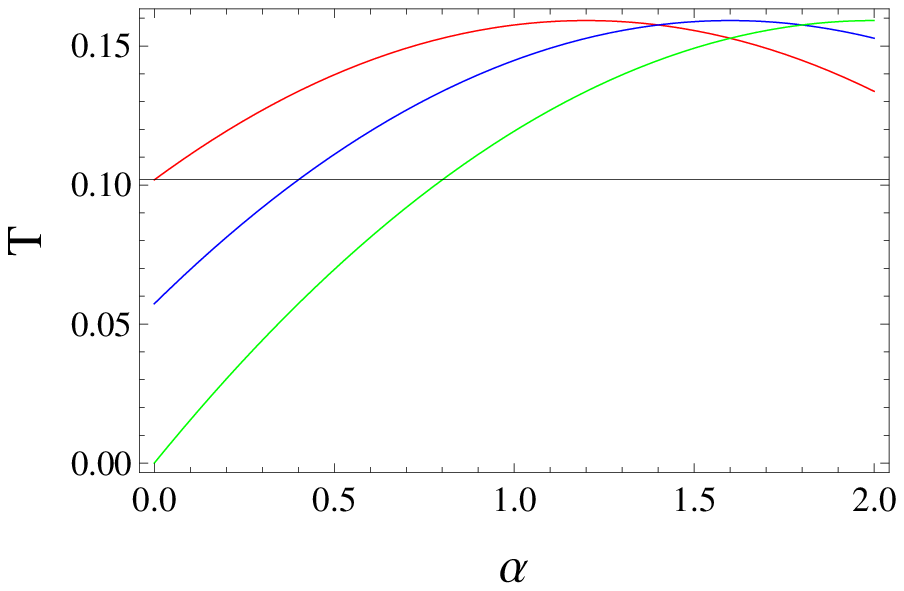,width=0.5\linewidth}\epsfig{file=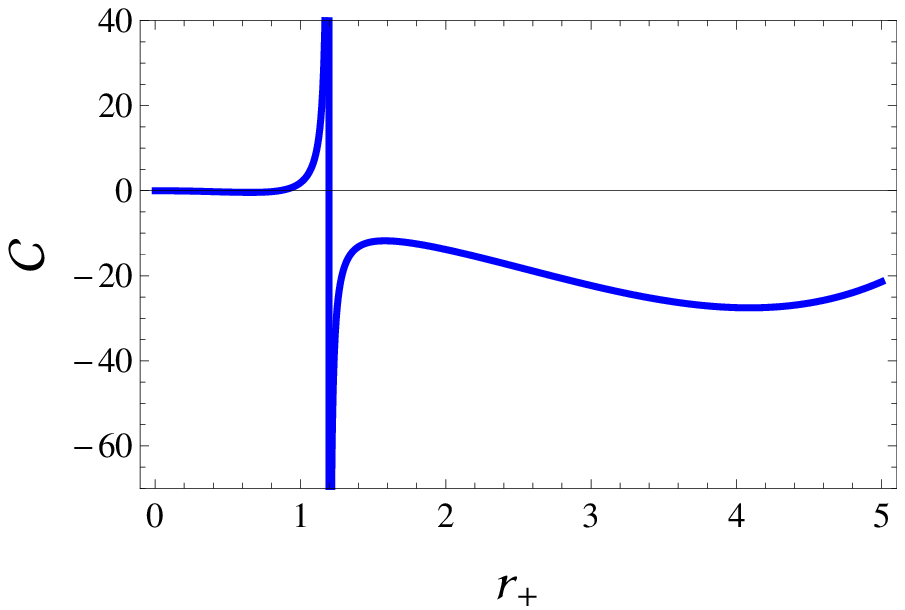,width=0.5\linewidth}
\caption{Plot of Hawking temperature (left) versus $\alpha$ for
$r_{+}=0.5$ with $q=0.3$(red), 0.4(blue), 0.5(green) and plot of
heat capacity (right) versus $r_{+}$ with $q=1$ and $\alpha=0.2$.}
\end{figure}

According to Bekenstein area-entropy relation, the entropy of BH can
be evaluated as \cite{33}
\begin{equation}\label{4}
S=\int_{0}^{2\pi}\int_{0}^{\pi}\sqrt{g_{\theta\theta}g_{\phi\phi}} d
\theta d \phi=\pi r_{+}^2.
\end{equation}
In order to analyze the thermal stability of BH, heat capacity
$(T\frac{\partial S}{\partial T})$ can be calculated as follows
\begin{equation}\label{4a}
C=\frac{\pi  r_{+}^2 \left(q-\alpha  r_{+}^2+r_{+}\right) (q-r_{+}
(\alpha r_{+}+1))}{-2 q^2+2 \alpha  q r_{+}^2+r_{+}^2}.
\end{equation}
It is noted that heat capacity of the considered system diverges at
$r_{+}=1.2$ (right plot of Figure \textbf{2}) and this divergent
point of heat capacity is known as Davies point. The positive region
before Davies point shows that BHs with small radii are stable while
negative region after Davies point shows that BHs with large radii
are unstable in the presence of $\alpha$.

Now we analyze the relationship between Davies point and QNMs. In
this regard, we calculate the heat capacity in terms of mass $M$.
For this purpose, the Hawking temperature in terms of $M$ is given
as
\begin{equation}\label{5}
T=\frac{3 M r_{+}-3 q^2-\alpha ^2 r_{+}^4}{6 \pi  r_{+}^3}.
\end{equation}
Consequently, the heat capacity becomes
\begin{equation}\label{6}
C=\frac{2 \pi  r_{+}^2 \left(-3 M r_{+}+3 q^2+\alpha ^2
r_{+}^4\right)}{6 M r_{+}-9 q^2+\alpha ^2 r_{+}^4}.
\end{equation}
The divergence point of heat capacity can be calculated by
considering the denominator of Eq.(\ref{6}) equal to zero as
\begin{equation}\nonumber
6 M r_{+}-9 q^2+\alpha ^2 r_{+}^4=0.
\end{equation}
Here,  it is difficult to get divergence point of heat capacity
because $q$ cannot be obtained explicitly in terms of $M$. In this
regard, we plot heat capacity to get divergence point and to analyze
its physical behavior.
\begin{figure}\center
\epsfig{file=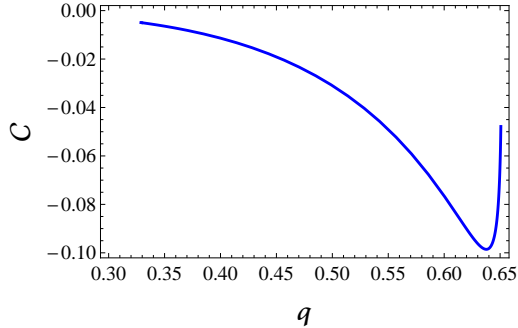,width=0.5\linewidth}\caption{Heat capacity
versus $q$ for $M=1$ and $\alpha=1.1$.}
\end{figure}
Figure \textbf{3} represents the graphical analysis of heat capacity
in terms of $q$ and shows that considered BH is thermodynamically
unstable. It is also noted that heat capacity diverges at $q=0.65$.

\section{Null Geodesics and Quasi-normal Modes}

This section is devoted to discuss the null geodesics and photon
sphere radius for RN BH with non-linear electrodynamics. We use
photon sphere radius to calculate angular velocity of the photon and
Lyapunov exponent. We restrict ourselves to the equatorial plane
($\theta=0,\frac{\pi}{2}$) and the corresponding Lagrangian becomes
\cite{34}
\begin{equation}\label{7}
2\mathcal{L}=\textit{f(r)} \dot{t}^{2} - \textit{f(r)}^{-1}
\dot{r}^{2}- {r}^2 \dot{\phi}^{2},
\end{equation}
where $\phi$ represents angular coordinate. The components
($\mathcal{P}_{u}=g_{uv}\dot{x}^{v}=\frac{\partial\mathcal{L}}{\partial
\dot{x}^{u}}$) of generalized momenta are given as
\begin{eqnarray}\label{8}
\mathcal{P}_{t}&=&f(r)\dot{t}=\bar{E}=constant,\\\label{9}
\mathcal{P}_{r}&=&-{f(r)}^{-1}\dot{r},\\\label{10}
\mathcal{P}_{\phi}&=&-{r}^2\dot{\phi}\equiv-l=constant,
\end{eqnarray}
where energy and angular momentum of the photon are represented by
conservation constants $\bar{E}$ and $l$, respectively. Using
Eqs.(\ref{8}) and (\ref{10}), $t$ and $\phi$-motions are evaluated
as
\begin{equation}\nonumber
\dot{t}={f(r)}^{-1}\bar{E},\quad \dot{\phi}=\frac{1}{{r}^2}l.
\end{equation}
The corresponding Hamiltonian yields
\begin{equation}\label{11}
2\mathcal{H}=-\dot{r}^{2}{f(r)}^{-1}+f(r)\dot{t}^{2}-{r}^2\dot{\phi}^{2}
=\bar{E}\dot{t}-\dot{r}^{2}{f(r)}^{-1}-l\dot{\phi}=0,
\end{equation}
which leads to
\begin{equation}\nonumber
V_{\text{eff}}=-\dot{r}^{2},
\end{equation}
where
\begin{equation}\label{12}
V_{\text{eff}}=\frac{f(r)l^{2}-\bar{E}^{2}r^{2}}{r^{2}}.
\end{equation}
This relation expresses that for positive $\dot{r}^{2}$, effective
potential becomes negative which indicates that photon cannot escape
from the region of negative potential. For large $l$ and small $r$,
the photon will drive out before falling inside the BH while for
large $r$ and small $l$, the photon will fall into the BH. However,
there is another region at a distance equal to the radius of BH
horizon where photon revolves around BH with zero radial velocity
\cite{34}. These circular orbits are unstable and are known as
photon sphere.

For spherically symmetric metric, photon sphere can be determined by
the following conditions
\begin{equation}\label{13}
V_{\text{eff}}=0,\quad \frac{\partial V_{\text{eff}}}{\partial
r}=0,\quad \frac{\partial^{2} V_{\text{eff}}}{\partial r^{2}}<0.
\end{equation}
Photon sphere radius $(r_{ps})$ can be evaluated by using first
condition of Eq.(\ref{13}) while the third condition ensures about
instability of photon sphere and associates with the QNMs of BH. By
using Eq.(\ref{12}) in the second condition, we obtain
\begin{equation}\label{14}
f^{'}(r_{ps})r_{ps}-2f(r_{ps})=0.
\end{equation}
For the given metric function (\ref{2}), it becomes
\begin{equation}\nonumber
3 M r_{ps} = 2 q^2 + r_{ps}^2 + 2 q r_{ps}^2 \alpha.
\end{equation}
The photon sphere radius is given as
\begin{equation}\label{15}
r_{ps}=\frac{3 M + \sqrt{9 M^2 - 8 q^2 - 16 q^3 \alpha}}{2 \left(1 +
2 q \alpha\right)}.
\end{equation}
This reduces to the photon sphere radius of RN BH for $\alpha=0$
\cite{11c}. In the eikonal limit $(l\gg1)$, QNMs can be evaluated by
using the property of the photon sphere \cite{35}
\begin{equation}\label{16}
w_{Q}=\Omega l-i|\lambda|\Big(\frac{2n+1}{2}\Big),
\end{equation}
where the number of overtune of the perturbations and angular
momentum of the photon are represented by $n$ and $l$, respectively.
The angular velocity $\Omega$ and Lyapunov exponent $\lambda$ are
two important quantities of the photon sphere. These are related to
QNMs as
\begin{equation}\label{17}
\Omega=\dot{\phi}\frac{1}{\dot{t}}\Big|_{r_{ps}}=\frac{\sqrt{f_{ps}}}{r_{ps}}
,\quad
\lambda=\sqrt{\frac{-V_{\text{eff}}^{''}}{2\dot{t}^{2}}}\Big|_{r_{ps}}
=\sqrt{\frac{(2f_{ps}-f_{ps}^{''}r_{ps}^{2})f_{ps}}{2r_{ps}^{2}}},
\end{equation}
which yield
\begin{eqnarray}\nonumber
\Omega&=&\frac{\sqrt{1 + \frac{q^2 - 2 M r_{ps}}{r_{ps}^2} + 2 q
\alpha - \frac{r_{ps}^2\alpha^2}{3}}}{r_{ps}},\\\nonumber \lambda
&=&\left(-\left(\left(-2 q^2 + r_{ps}^2 + 2 q r_{ps}^2 \alpha\right)
\left(-3 \left(q^2 + r_{ps} \left(-2 M + r_{ps}\right)\right)
\right.\right.\right.\\\nonumber&-&\left.\left.\left. 6 q r_{ps}^2
\alpha + r_{ps}^4 \alpha^2\right)\right)\left(3
r_{ps}^6\right)\right)^{\frac{1}{2}}.
\end{eqnarray}

The graphical behavior of angular velocity versus $q$ shows that
there is no effect of coupling parameter on the angular velocity of
photon for the considered values (Figure \textbf{4} (left)). It is
also observed that angular velocity of photon is not defined before
$q=0.65$ which is the Davies point of $\Omega$. However, the right
plot shows that increasing rate of angular velocity and radius of
photon sphere are inversely proportional to each other. It is
interesting to mention here that the Davies point noted from Figures
\textbf{3} and \textbf{4} are exactly the same. Figure \textbf{5}
shows the graphical representation of lyapunov exponent versus $q$.
It is observed that lyapunov exponent shows increasing behavior for
increasing values of photon sphere radius and coupling parameters
which indicate the increasing rate of modes of perturbation.
\begin{figure}\center
\epsfig{file=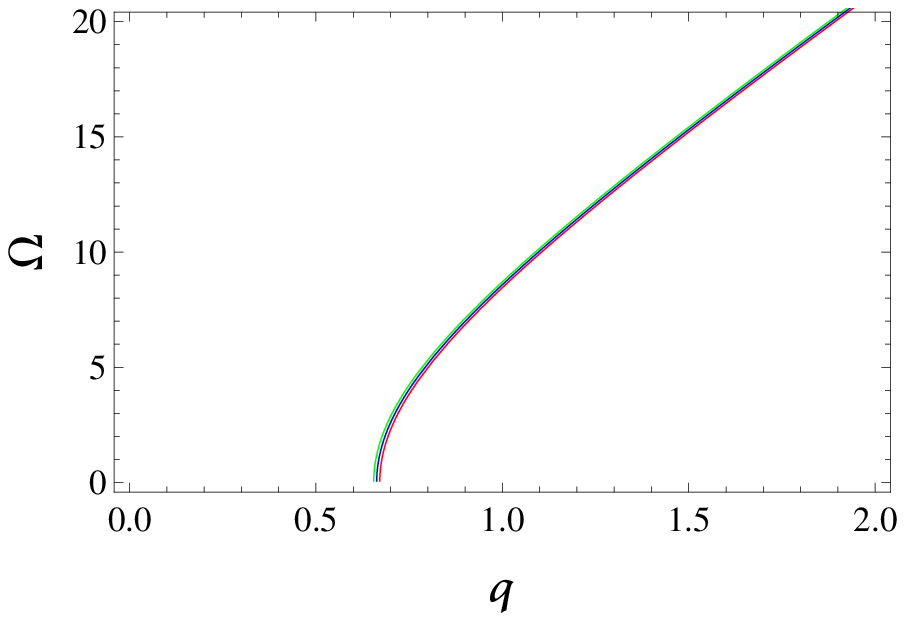,width=0.5\linewidth}\epsfig{file=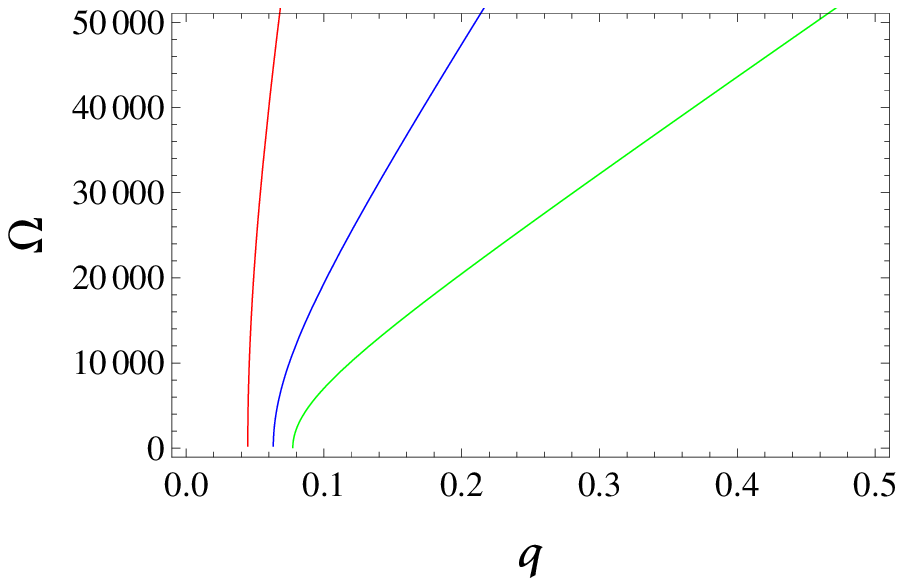,width=0.5\linewidth}\caption{Plots
of $\Omega$ versus $q$ for $M=1$, $r_{ps}=0.3$ and
$\alpha=0.5$(red), 0.6(blue), 0.7(green) for the left plot and
$M=\alpha=1$ with $r_{ps}=0.001$(red), 0.002(blue), 0.003(green) for
the right plot.}
\end{figure}
\begin{figure}\center
\epsfig{file=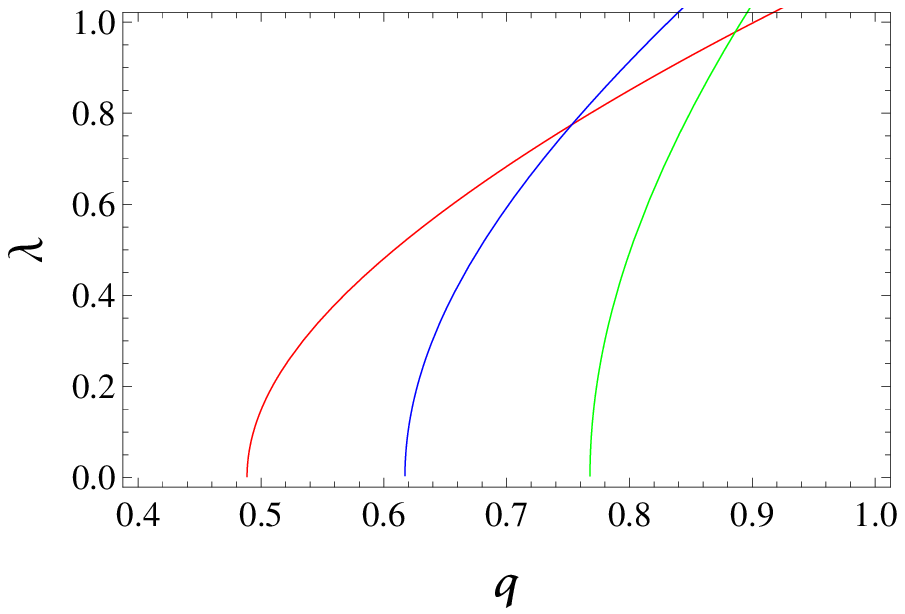,width=0.5\linewidth}\epsfig{file=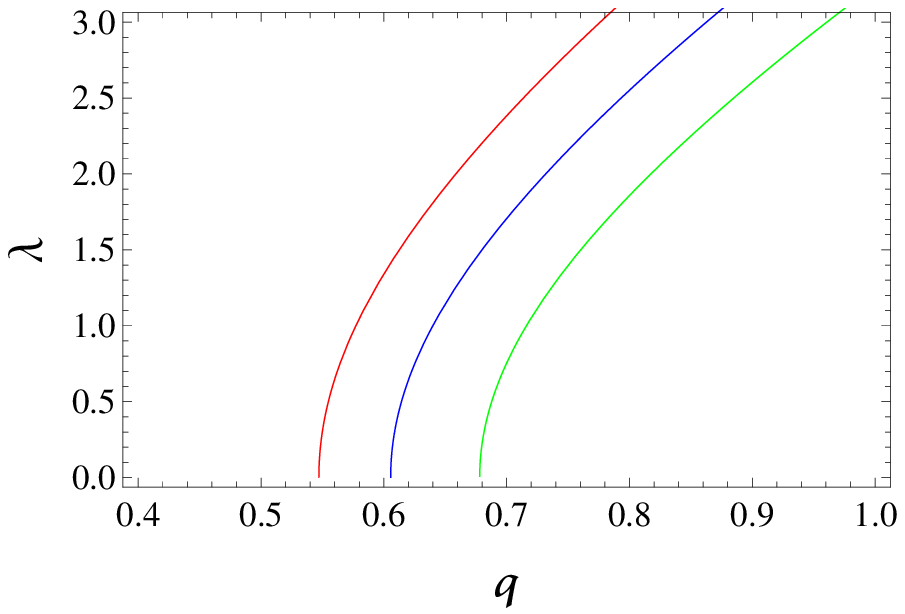,width=0.5\linewidth}
\caption{Plots of $\lambda$ versus $q$ for $M=1$, $r_{ps}=1.5$ with
$\alpha=1$(red), 1.5(blue), 2(green) for the left plot and $M=1$,
$\alpha=3$ with $r_{ps}=0.9$(red), 1(blue), 1.1(green) for the right
plot.}
\end{figure}

\section{Thermal Fluctuations}

In this section, we investigate the effects of thermal fluctuations
on thermodynamical potentials of RN BH with non-linear
electrodynamics. We compute the corrected and uncorrected
expressions of physical quantities. By using Eq.(\ref{2}) and taking
$f(r_{+})=0$, we have
\begin{equation}\label{18}
3 r_{+}^2-6r_{+}M+3q^2+6r_{+}q\alpha-\alpha^2 r_{+}^4=0.
\end{equation}
Consequently, the Hawking temperature turns out to be
\begin{equation}\nonumber
T=\frac{3 M r_{+}-3 q^2-\alpha ^2 r_{+}^4}{6 \pi  r_{+}^3}.
\end{equation}
In order to evaluate corrected expression for entropy, we define the
partition function as \cite{16a}
\begin{equation}\label{19}
R(\xi)=\int_{0}^{\infty} \exp(-\xi E)\rho(E)dE ,
\end{equation}
where $\rho(E)$ and $E$ are the density of state and average energy,
respectively. The inverse Laplace transform of the above partition
function gives the density of state in the following form
\begin{equation}\label{20}
\rho(E)=\frac{1}{2i\pi
}\int_{-i\infty+\xi_{0}}^{i\infty+\xi_{0}}\exp(\xi E) R(\xi) d\xi
=\frac{1}{2i\pi }\int_{-i\infty+\xi_{0}}^{i\infty+\xi_{0}}
\exp(\tilde{S}(\xi))d\xi,
\end{equation}
where $\tilde{S}(\xi)=\beta E+\ln Z(\xi)$ with $\xi>0$ is referred
to as exact expression for entropy of the BH and has dependence on
Hawking temperature. Using Taylor series expansion, we have
\begin{equation}\label{21}
\tilde{S}(\xi)=S+\frac{1}{2}(\xi-\xi_{0})^{2}
\frac{\partial^{2}\tilde{S}(\xi)}{\partial
\xi^{2}}\Big|_{\xi=\xi_{0}}+ O(\xi-\xi_{o})^{2}.
\end{equation}
The equilibrium entropy $S$ satisfies the relations $\frac{\partial
S}{\partial\xi}=0$ and $\frac{\partial^{2} S}{\partial\xi^{2}}>0$.
By using Eq.(\ref{21}) in (\ref{20}), we obtain
\begin{equation}\label{22}
\rho(E)=\frac{1}{2\pi i}\exp(S)\int d\xi
\exp\Big(\frac{1}{2}(\xi-\xi_{0})^{2}
\frac{\partial^{2}\tilde{S}(\xi)}{\partial \xi^{2}}\Big).
\end{equation}
Further, it can be written as \cite{37}
\begin{equation}\label{23}
\rho(E)=\frac{1}{\sqrt{2\pi}}\exp(S)\Big(\Big(\frac{\partial^{2}\tilde{S}(\xi)
}{\partial \xi^{2}}\Big)\Big|_{\xi=\xi_{0}}\Big)^{-\frac{1}{2}},
\end{equation}
which yields
\begin{equation}\label{24}
\tilde{S}=S-\frac{1}{2}\ln(ST^{2})+\frac{\eta}{S}.
\end{equation}
Without loss of generality, we can use a more general parameter
$\gamma$ except the factor $\frac{1}{2}$ which increases the effect
of correction terms on the entropy of BH. In this context, the
corrected entropy can be written as \cite{38}
\begin{equation}\label{25}
\tilde{S}=S-\gamma \ln(ST^{2})+\frac{\eta}{S}.
\end{equation}
For different values of correction parameters $\gamma$ and $\eta$,
we obtain
\begin{itemize}
\item For $\eta,~ \gamma \rightarrow 0$, we have the uncorrected entropy
of BH.
\item When $\eta \rightarrow 0,~ \gamma \rightarrow 1$,
this provides simple logarithmic corrections.
\item When $\eta \rightarrow 1,~ \gamma \rightarrow 0$, we obtain
second order correction terms which are inversely proportional to
the uncorrected entropy of the BH.
\item For $\eta ,~ \gamma\rightarrow 1$, we obtain correction terms of higher orders.
\end{itemize}

In the following, we only consider the second case ($\eta\rightarrow
0, \gamma \rightarrow 1$). The second term in Eq.(\ref{25}) is
logarithmic which leads to the corrections in entropy. The perturbed
expression of entropy can be obtained by using Eqs.(\ref{4}) and
(\ref{5}) in (\ref{25}) as
\begin{equation}\label{26}
\tilde{S}=\pi  r_{+}^2-2 \gamma  \ln \left(3 M r_{+}-3 q^2-\alpha ^2
r_{+}^4\right)+\gamma  \ln \left(36 \pi  r_{+}^4\right).
\end{equation}
Figure \textbf{6} (left plots) represents the graphical behavior of
corrected entropy in terms of horizon radius for different values of
$\alpha$, $\gamma$ and $q$, respectively. It is noted that corrected
entropy remains positive for considered values of $\alpha$, $\gamma$
as well as $q$ and hence satisfies the second law of thermodynamics.
The first and third plots represent that thermal fluctuations and
charge only affect the entropy of BHs negligibly for small radii
while for large radii it remains in equilibrium position. However,
the parameter $\alpha$ equally affects the entropy of both BHs with
small as well as large radii (2nd plot).
\begin{figure}\center
\epsfig{file=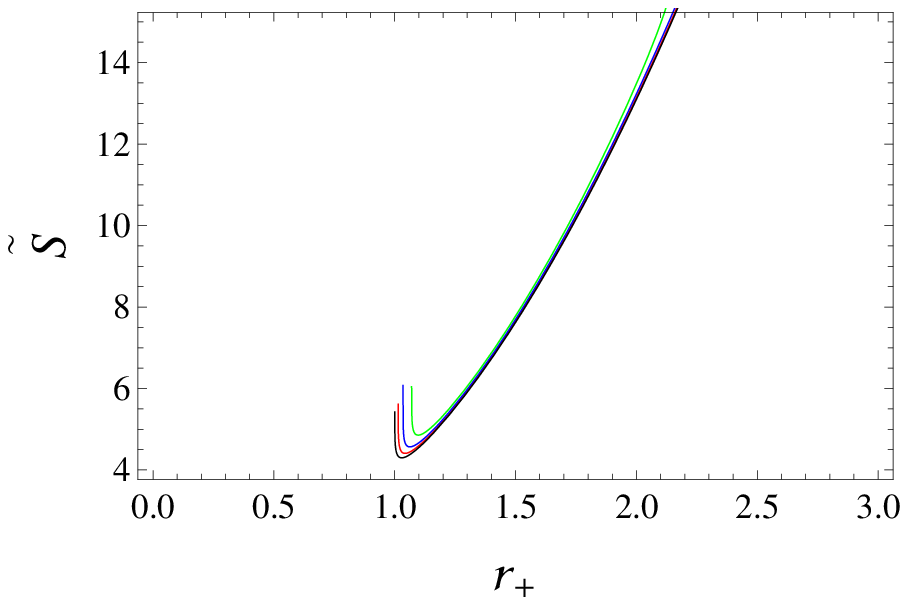,width=0.5\linewidth}\epsfig{file=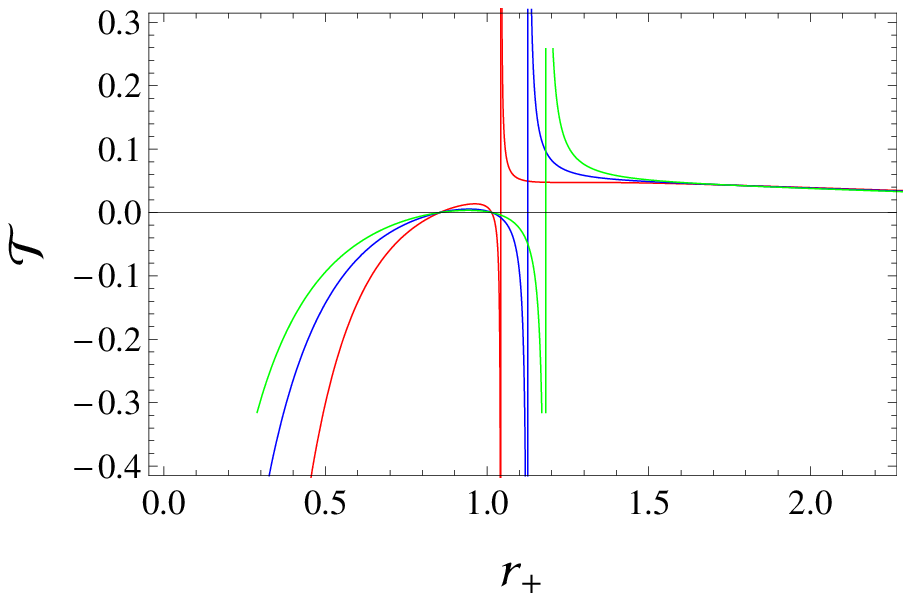,width=0.5\linewidth}
\epsfig{file=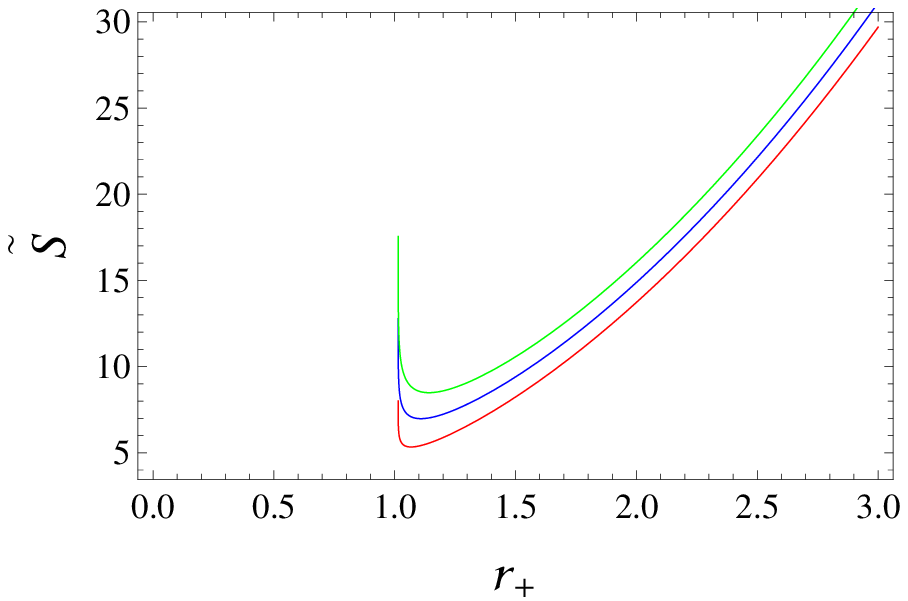,width=0.5\linewidth}\epsfig{file=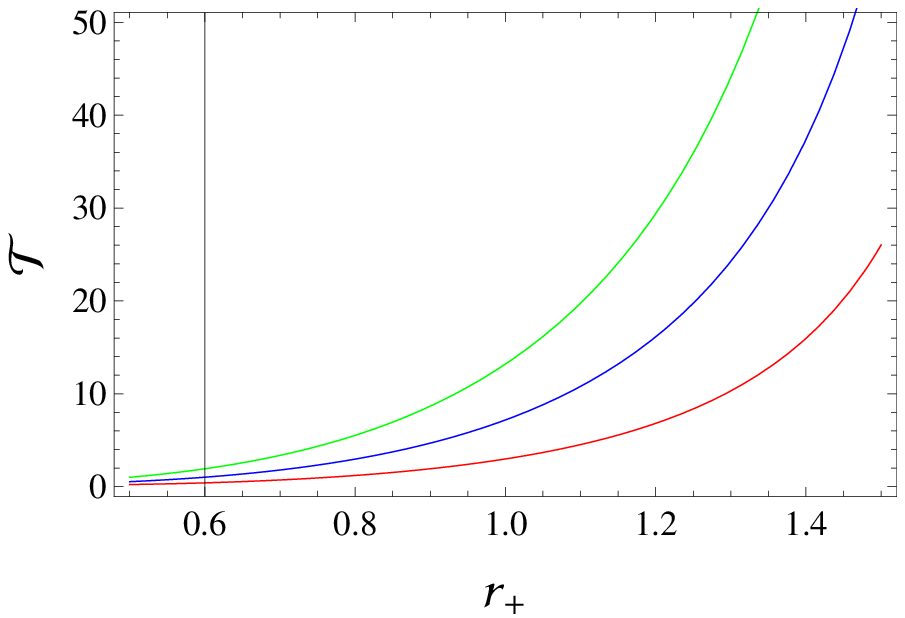,width=0.5\linewidth}
\epsfig{file=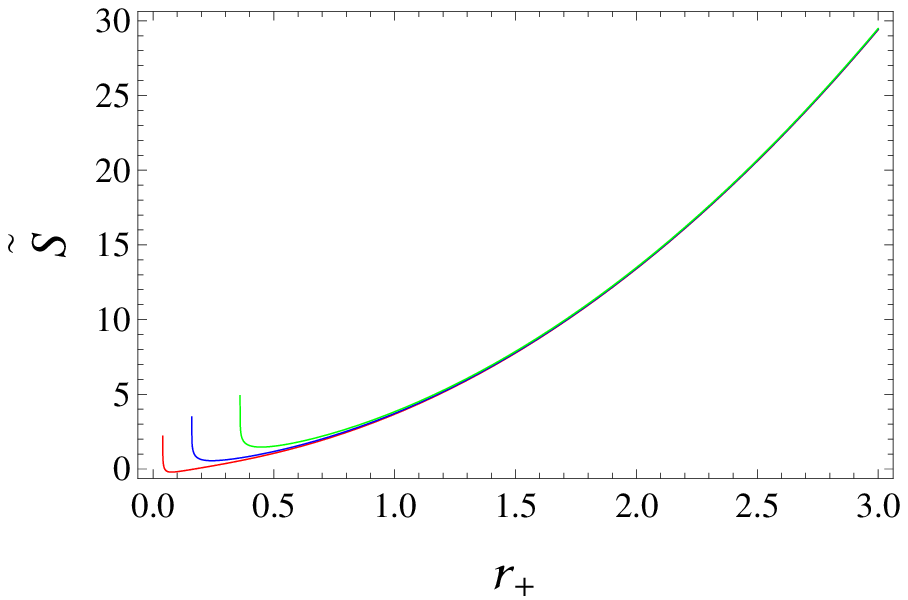,width=0.5\linewidth}\epsfig{file=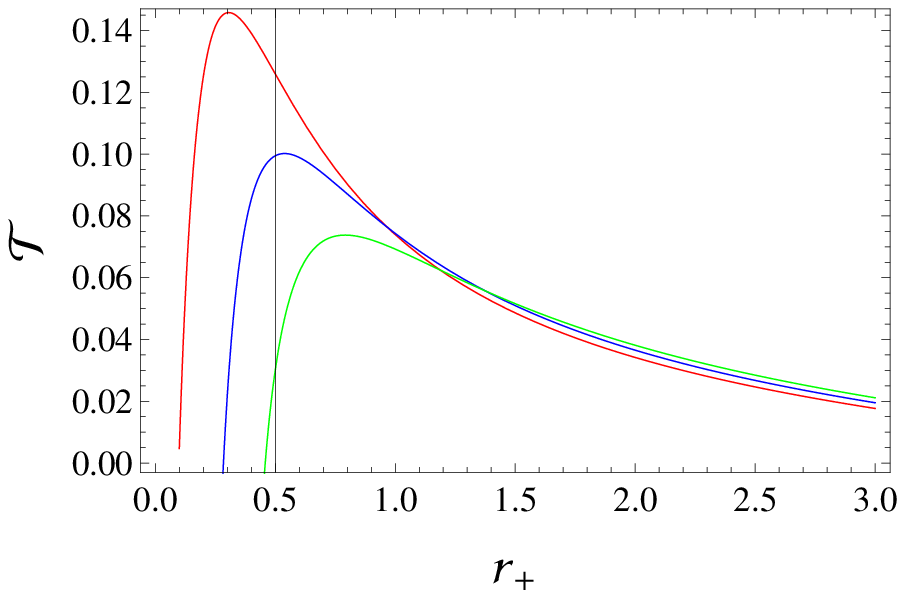,width=0.5\linewidth}\caption{Corrected
entropy (left 3 plots) versus $r_{+}$ for $M=q=1$. We take
$\gamma=0.1$, with $\alpha=0$(black), 0.2(red), 0.3(blue),
0.4(green) for the 1st plot, $\alpha=0.2$ with $\gamma=0.2$(red),
0.4(blue), 0.6(green) for the 2nd plot and $M=1$,
$\alpha=\gamma=0.2$ with $q=0.2$(red), 0.4(blue), 0.6(green) for the
3rd plot. Corrected Hawking temperature versus $r_{+}$ with $M=q=1$,
$\alpha=0.2$ with $\gamma=0.1$(red), 0.5(blue), 0.9(green) for the
1st plot, $\gamma=5$ with $\alpha=10$(red), 15(blue), 20(green) for
the 2nd plot and $M=1$, $\alpha=\gamma=0.2$ with $q=0.1$(red),
0.3(blue), 0.5(green) for the 3rd plot.}
\end{figure}

The modified first law of BH thermodynamics under thermal
fluctuations can be written as \cite{15}
\begin{eqnarray}\label{27}
dM&=&\mathcal{T} dS+\varphi dq+\Pi d\alpha,,
\end{eqnarray}
where $\mathcal{T}$ and $\varphi$ are corrected Hawking temperature
and electric potential, respectively. Here, we consider $\alpha$ as
a new thermodynamic variable and $\Pi$ is the corresponding
conjugate quantity. These potential functions can be obtained from
the relations
\begin{equation}\nonumber
\mathcal{T}=\Big(\frac{\partial M}{\partial
\tilde{S}}\Big)_{q,\alpha},\quad \varphi=\Big(\frac{\partial
M}{\partial q}\Big)_{T,\alpha},\quad \Pi=\Big(\frac{\partial
M}{\partial \alpha}\Big)_{T,q}
\end{equation}
which yield
\begin{eqnarray}\nonumber
\mathcal{T}&=&\left(1-\frac{\left(q-\alpha
r_{+}^2\right)^2}{r_{+}^2}\right)\left(2 \left(\frac{6 \gamma M-8
\alpha ^2 \gamma  r_{+}^3}{-3 M r_{+}+3 q^2+\alpha ^2
r_{+}^4}+\frac{4 \gamma }{r_{+}}+2 \pi
r_{+}\right)\right)^{-1},\\\nonumber \varphi&=&\frac{3}{2}
\left(1-\frac{\left(q-\alpha  r_{+}^2\right)^2}{r_{+}^2}\right)
\left(\frac{\sqrt{3} \left(3 M+8 \alpha ^2 r_{+}^3-3
r_{+}\right)}{\sqrt{r_{+} \left(6 M+4 \alpha ^2 r_{+}^3-3
r_{+}\right)}}-6 \alpha r_{+}\right)^{-1},\\\nonumber \Pi&=&\frac{-6
q \sqrt{r_{+}^4 \left(-2 M r_{+}+4 q^2+r_{+}^2\right)}+\sqrt{3}
r_{+}^3 (3 M-r_{+})-8 \sqrt{3} q^2 r_{+}^2}{r_{+}^3 \sqrt{r_{+}^4
\left(r_{+} (r_{+}-2 M)+4 q^2\right)}}.
\end{eqnarray}
The modified first law of BH gets satisfied on substitution of above
expressions in Eq.(\ref{27}) indicating that the presence of thermal
fluctuations increases the validity of first law of BH
thermodynamics.

The graphical analysis of corrected Hawking temperature versus
$r_{+}$ is shown in Figure \textbf{6} (right plots). It is observed
that temperature of BHs with small radii are affected by thermal
fluctuations while the region after the divergence point shows that
BHs of large radii are unaffected (first plot). It is also observed
that critical radius of BH increases by increasing the values of
correction parameter. The second plot shows the profile of corrected
Hawking temperature for different values of coupling parameter. We
note that the corrected Hawking temperature increases monotonically
with respect to horizon radius. However, with increasing values of
charge, temperature firstly increases and then shows decreasing
behavior for large radii (3rd plot).

Now, we use the expressions of corrected entropy as well as Hawking
temperature to explore thermodynamical equations of state. In this
regard, the Helmholtz free energy can be obtained by using the
relation
\begin{equation}\label{28}
F=-\int \tilde{S} dT.
\end{equation}
Inserting the expressions of $\tilde{S}$ and $T$, the Helmholtz free
energy is given as
\begin{eqnarray}\nonumber
F&=&\left(3 \gamma  \left(-3 M r_{+}+4 q^2+4 \alpha ^2
r_{+}^4\right)+3 \gamma \left(-3 M r_{+}+3 q^2+\alpha ^2
r_{+}^4\right)\right.\\\nonumber&\times&\left. \left(\ln \left(36
\pi r_{+}^4\right)-2 \ln \left(3 M r_{+}-3 q^2-\alpha ^2
r_{+}^4\right)\right)+18 \pi M r_{+}^3 \ln
(r_{+})\right.\\\nonumber&+&\left.\pi \left(27 q^2 r_{+}^2+\alpha ^2
r_{+}^6\right)\right)\left(18 \pi r_{+}^3\right)^{-1}.
\end{eqnarray}
The graphical representation of Helmholtz free energy versus $r_{+}$
is shown in Figure \textbf{7} (left plots). It is observed that the
Helmholtz free energy remains positive throughout the considered
domain for increasing values of $\gamma$ and $\alpha$ (1st and 2nd
plots). Thermal fluctuations affect the Helmholtz free energy of BHs
with small radii more as compared to the BHs of large radii for
increasing values of $\gamma$ while for different values of
$\alpha$, it shows opposite trend. However, the third plot
represents that Helmholtz free energy increases from negative to
positive values with increasing values of $q$.
\begin{figure}\center
\epsfig{file=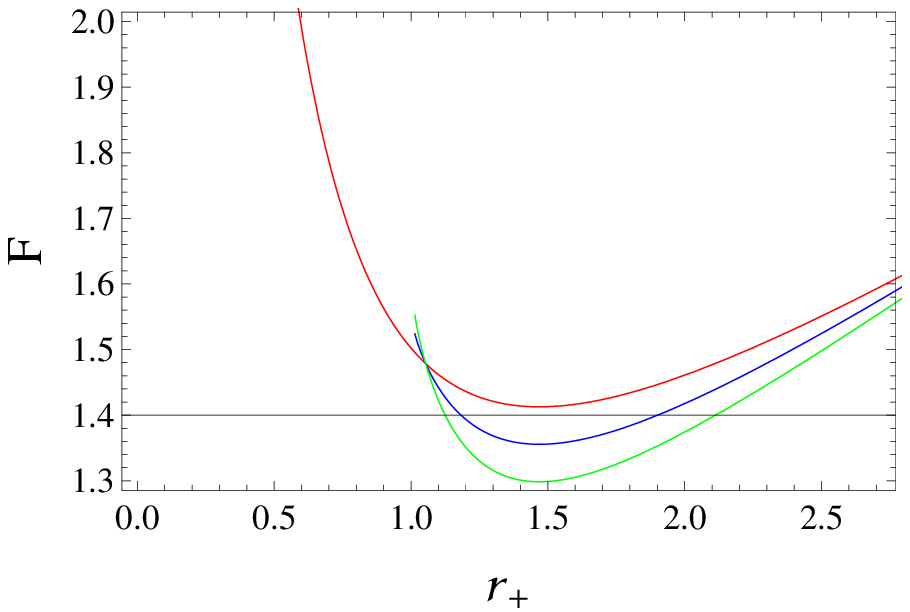,width=0.5\linewidth}\epsfig{file=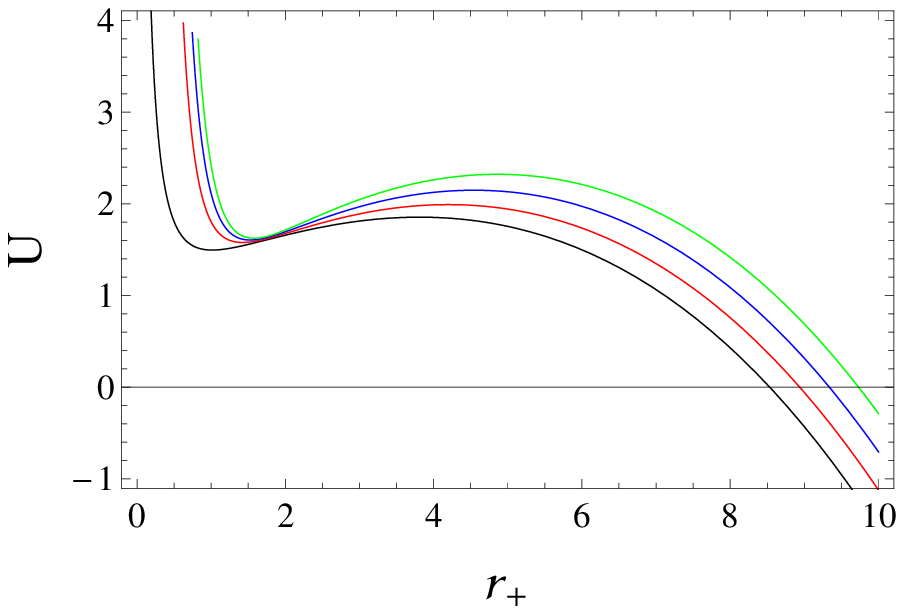,width=0.5\linewidth}
\epsfig{file=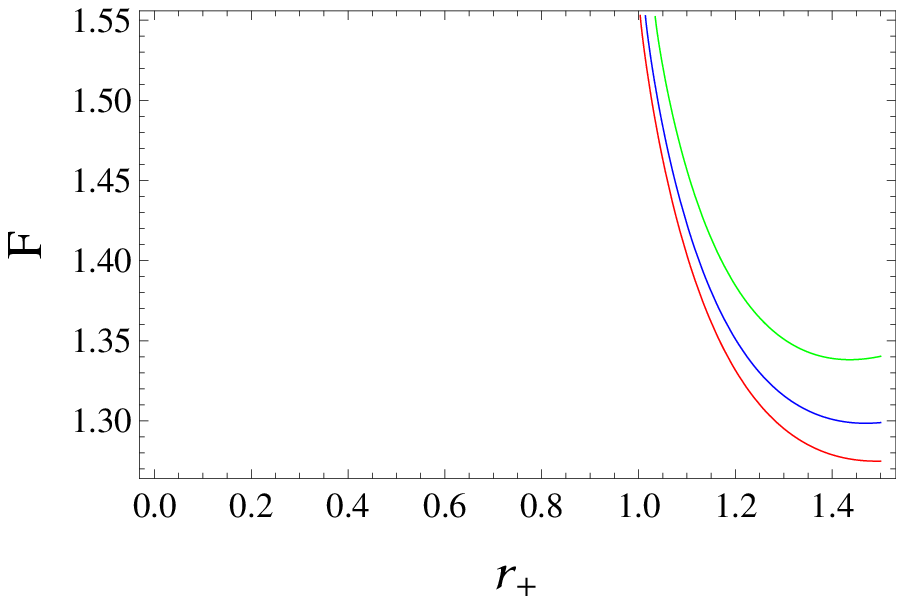,width=0.5\linewidth}\epsfig{file=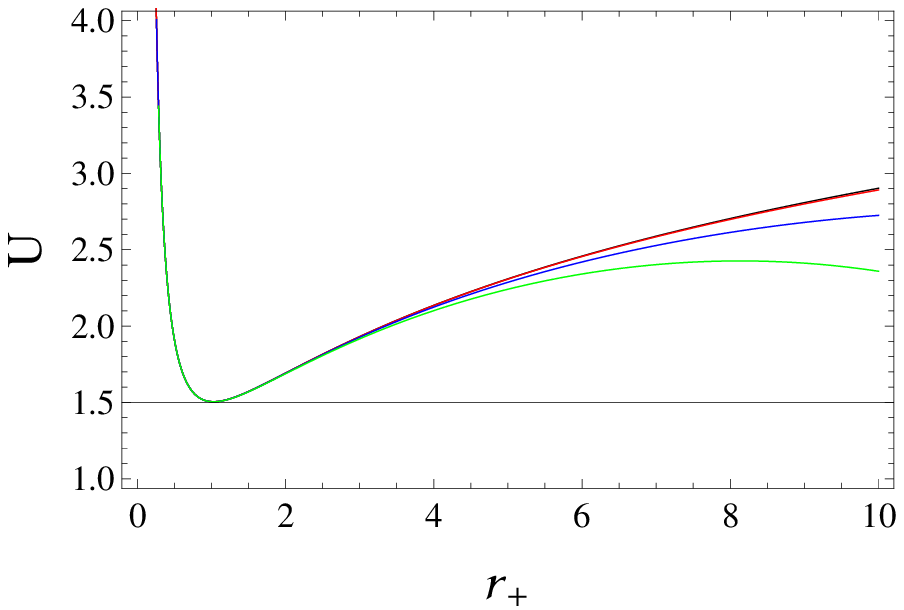,width=0.5\linewidth}
\epsfig{file=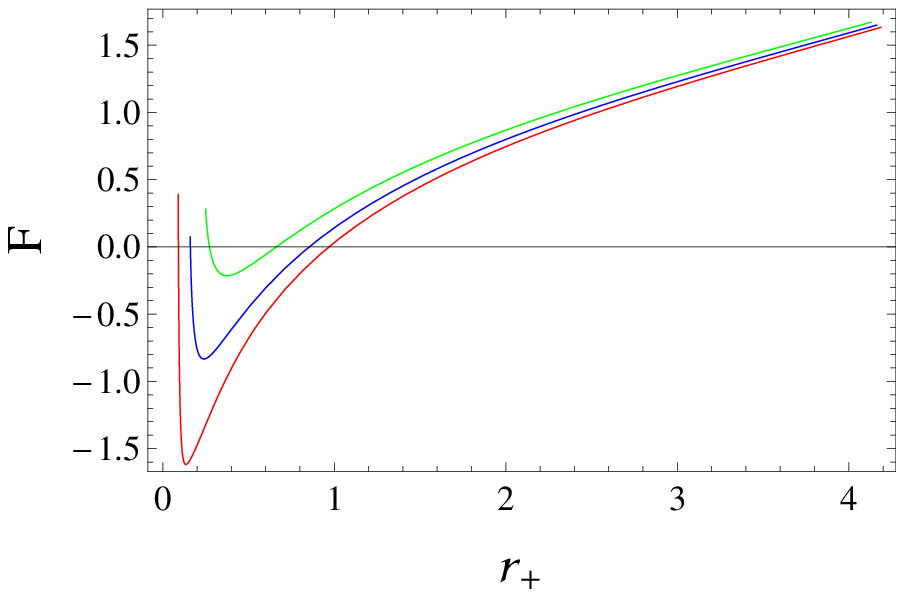,width=0.5\linewidth}\epsfig{file=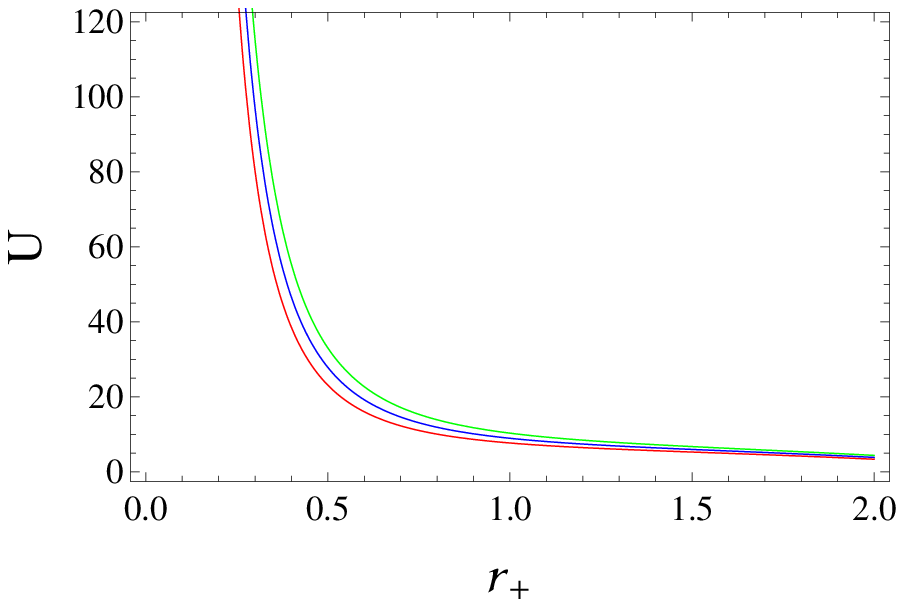,width=0.5\linewidth}
\caption{Helmholtz free energy versus $r_{+}$ for $M=q=1$. We take
$\alpha=0.2$ with $\gamma=0$(red), 0.5(blue), 1(green) for the 1st
plot, $\gamma=1$ with $\alpha=0.1$(red), 0.2(blue), 0.3(green) for
the 2nd plot and $M=1$, $\alpha=\gamma=0.2$ with $q=0.3$(red),
0.4(blue). 0.5(green) for the 3rd plot. Internal energy versus
$r_{+}$ for $M=g=1$. We take $\alpha=0.2$ with $\gamma=0$(black),
5(red), 10(blue), 15(green) for the 1st plot, $\gamma=0.1$ with
$\alpha=0$(black), 0.01(red), 0.04(blue), 0.07(green) for the 2nd
plot and $M=1$, $\alpha=\gamma=2$ with $q=2.1$(red), 2.3(blue),
2.5(green) for the 3rd plot.}
\end{figure}

The internal energy of the considered system can be evaluated by the
identity, $U=\tilde{S}T+F$ \cite{35} as
\begin{eqnarray}\nonumber
U&=&\left(3 \gamma  \left(-3 M r_{+}+4 q^2+4 \alpha ^2
r_{+}^4\right)+\pi r_{+}^2 \left(9 M r_{+}+18 q^2-2 \alpha ^2
r_{+}^4\right)\right.\\\label{29}&+&\left.18 \pi M r_{+}^3 \ln
(r_{+})\right)\left(18 \pi  r_{+}^3\right)^{-1}.
\end{eqnarray}
Figure \textbf{7} (right plots) shows the graphical analysis of
internal energy versus $r_{+}$ for different values of the
correction and coupling parameters. It is noted that the internal
energy fluctuates with the increasing values of correction parameter
and becomes negative for large values of horizon radius (1st plot).
The left plot shows that internal energy decreases and remains
unaffected by thermal fluctuations for small values of $r_{+}$ while
it shows increasing behavior for large values of $\alpha$. However,
third plot represents that BHs with large values of charge have less
internal energy and vice versa.

The volume of the BH can be given as \cite{38}
\begin{equation}\label{30}
V=\frac{4}{3}\pi r_{+}^{3}.
\end{equation}
The pressure of the BH can be calculated by the relation
\begin{equation}\label{31}
P=-\frac{dF}{dV}=-\frac{dF}{dr_{+}}\frac{dr_{+}}{dV},
\end{equation}
hence
\begin{eqnarray}\nonumber
P&=&-\left(\left(6 M r_{+}-9 q^2+\alpha ^2 r_{+}^4\right) \left(-2
\gamma \ln \left(3 M r_{+}-3 q^2-\alpha ^2
r_{+}^4\right)\right.\right.\\\nonumber&+&\left.\left.\gamma \ln
\left(36 \pi r_{+}^4\right)+\pi r_{+}^2\right)\right)\left(24 \pi ^2
r_{+}^6\right)^{-1}.
\end{eqnarray}
The graphical behavior of pressure versus $r_{+}$ for different
values of $\gamma$, $\alpha$ and $q$ is shown in Figure \textbf{8}
(left plots). It is observed that for $\gamma=0$ (no thermal
fluctuations) pressure firstly increases for small radii and becomes
negative and it increases continuously for large radii but remains
negative throughout the considered domain. However, in the presence
of thermal fluctuations, it is affected significantly and is not
defined for $r_{+}>3.8$ (1st plot). The second plot also shows that
pressure is decreasing and becomes negative for increasing values of
$\alpha$ while it becomes singular for $\alpha>0.556$. For
increasing values of charge pressure, it also increases and
coincides with equilibrium condition but remains negative for the
considered domain.
\begin{figure}\center
\epsfig{file=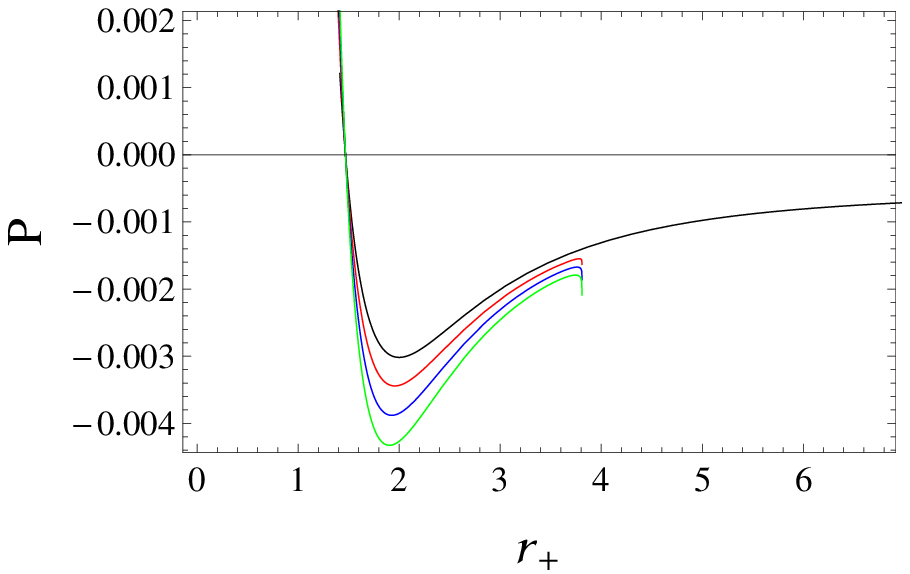,width=0.5\linewidth}\epsfig{file=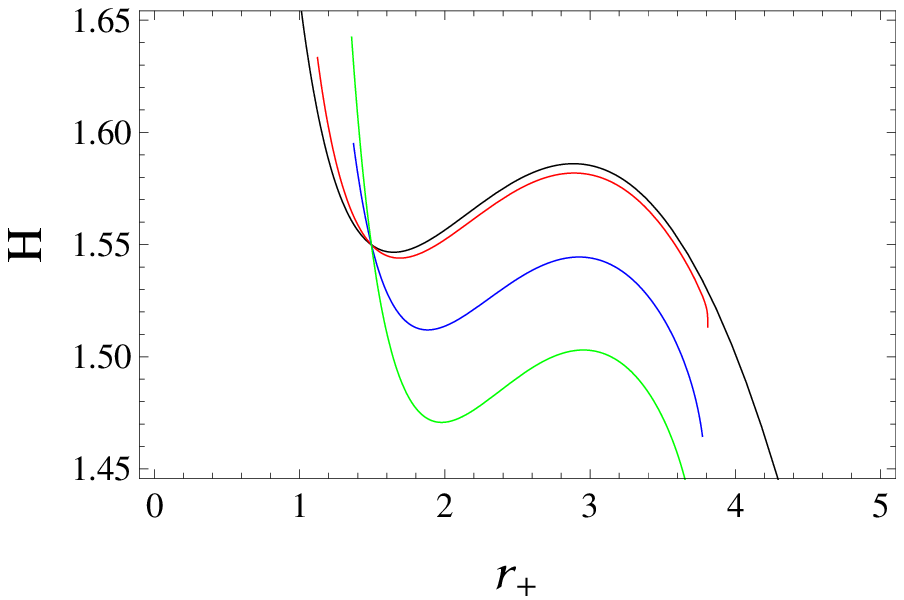,width=0.5\linewidth}
\epsfig{file=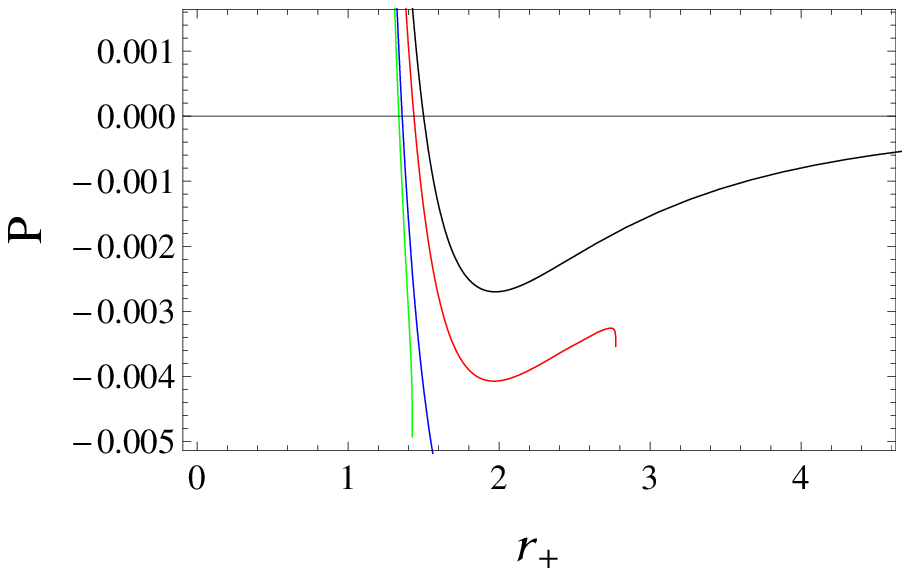,width=0.5\linewidth}\epsfig{file=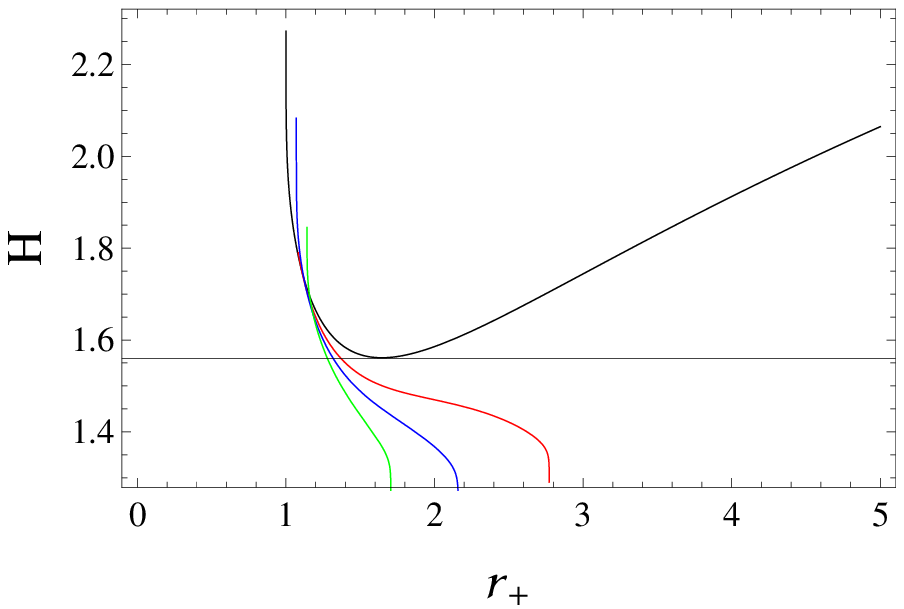,width=0.5\linewidth}
\epsfig{file=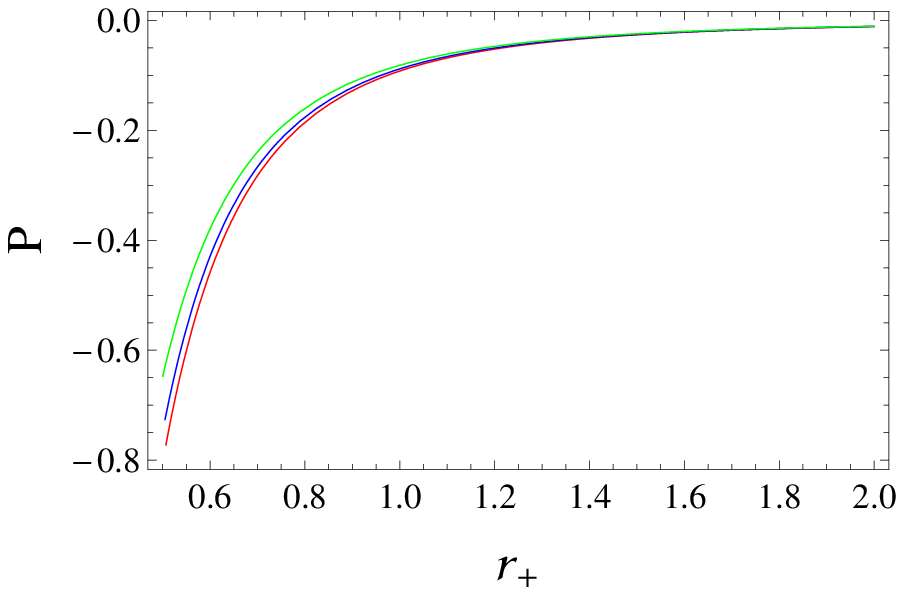,width=0.5\linewidth}\epsfig{file=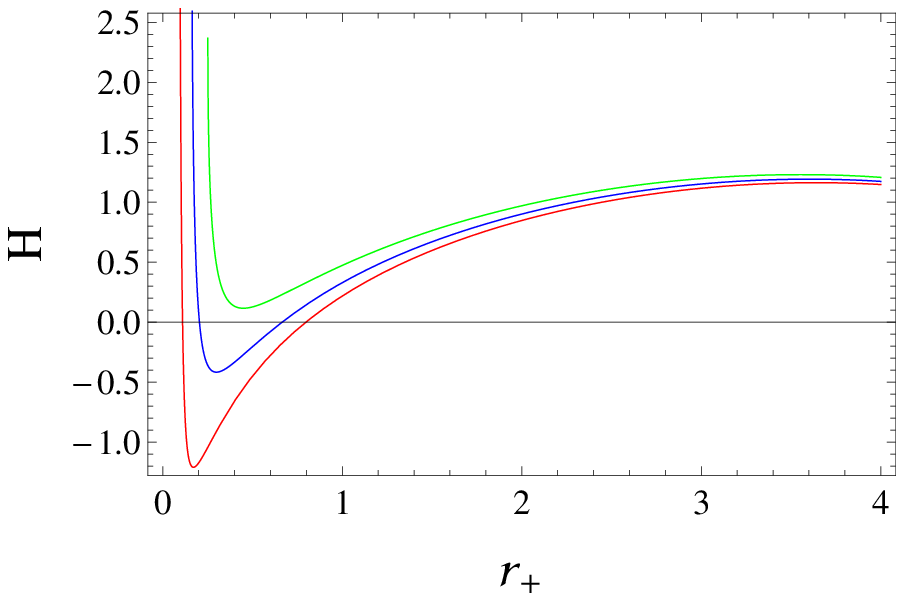,width=0.5\linewidth}
\caption{Pressure (left 3 plots) versus $r_{+}$ with $M=q=1$. We
take $\alpha=0.2$ with $\gamma=0$(black), 0.3(red), 0.6(blue),
0.9(green) for the 1st plot, $\gamma=0.2$ with $\alpha=0$(black),
0.3(red), 0.5(blue), 0.556(green) for the 2nd plot and $M=1$,
$\alpha=\gamma=0.2$ with $q=0.1$(red), 0.2(blue), 0.3(green) for the
3rd plot. Enthalpy (right 3 plots) versus $r_{+}$ for $M=q=1$. We
take $\alpha=0.2$ with $\gamma=0$(black), 0.1(red), 1(blue),
2(green) for the 1st plot, $\gamma=0.5$ with $\alpha=0$(black),
0.3(red), 0.4(blue), 0.5(green) for the 2nd plot and and $M=1$,
$\alpha=\gamma=0.2$ with $q=0.3$(red), 0.4(blue), 0.5(green) for the
3rd plot.}
\end{figure}

The enthalpy $(H=U+VP)$ of the system can be evaluated as
\begin{eqnarray}\nonumber
H&=&\left(3 \gamma  \left(-3 M r_{+}+4 q^2+4 \alpha ^2
r_{+}^4\right)+\left(-6 M r_{+}+9 q^2-\alpha ^2
r_{+}^4\right)\right.\\\nonumber&\times&\left. \left(-2 \gamma \ln
\left(3 M r_{+}-3 q^2-\alpha ^2 r_{+}^4\right)+\gamma \ln \left(36
\pi r_{+}^4\right)+\pi r_{+}^2\right)\right.\\\nonumber&+&\left.\pi
r_{+}^2 \left(9 M r_{+}+18 q^2-2 \alpha ^2 r_{+}^4\right)+18 \pi M
r_{+}^3 \ln (r_{+})\right)\left(18 \pi r_{+}^3\right)^{-1}.
\end{eqnarray}
Figure \textbf{8} (right plots) shows that enthalpy of the system
remains positive throughout the considered domain for different
values of $\gamma$ and $\alpha$. It is observed that for both with
and without thermal fluctuations, enthalpy of the considered system
fluctuates in terms of $r_{+}$ (1st plot). However, the second plot
shows that it decreases for small radii and then increases gradually
for BHs of large radii in the absence of non-linear electrodynamic
effects. For non-zero $\alpha$, enthalpy of the system decreases
gradually and is affected only for BHs of large radii. However, for
increasing values of charge enthalpy of the considered system also
increases and gets equilibrium position for large radii (3rd plot).

The Gibbs free energy $(G=-\tilde{S}T+H)$ can be obtained as
\begin{eqnarray}\nonumber
G&=&\left(3 \gamma  \left(-3 M r_{+}+4 q^2+4 \alpha ^2
r_{+}^4\right)+\gamma \left(-15 M r_{+}+18 q^2+2 \alpha ^2
r_{+}^4\right)\right.\\\nonumber&\times&\left. \left(\ln \left(36
\pi r_{+}^4\right)-2 \ln \left(3 M r_{+}-3 q^2-\alpha ^2
r_{+}^4\right)\right)-6 \pi  r_{+}^2 \left(M r_{+}-6
q^2\right)\right.\\\nonumber&+&\left.18 \pi  M r_{+}^3 \ln
(r_{+})\right)\left(18 \pi r_{+}^3\right)^{-1}.
\end{eqnarray}
Figure \textbf{9} (left plots) indicates the graphical behavior of
Gibbs free energy in terms of $r_{+}$ for different values of
$\gamma$, $\alpha$ and $q$. For $\gamma=0$ (no thermal
fluctuations), Gibbs free enrgy firstly decreases for small values
of $r_{+}$ but remains positive and increases for large values of
$r_{+}$. The presence of thermal fluctuations affects the Gibbs free
energy with both small as well as large radii and ultimately becomes
undefined for $r_{+}>3.8$ (1st plot). However, the second plot shows
that for $\alpha=0$, the Gibbs free energy firstly decreases and
becomes negative for BHs of small radii and then becomes positive by
increasing for BHs of large radii. In the presence of non-linear
electrodynamics, the Gibbs free energy of BHs with large radii are
affected notably and BHs of small radii are affected negligibly.
However, for different values of charge, the Gibbs free energy shows
increasing behavior and coincides for large values of horizon
radius.

We investigate thermal stability of RN BH with non-linear
electrodynamic effects by means of specific heat
$(C_{S}=\frac{dU}{dT})$ \cite{35} given as
\begin{eqnarray}\nonumber
C_{S}&=&-\frac{2 \left(\gamma  \left(3 M r_{+}-6 q^2+2 \alpha ^2
r_{+}^4\right)-\pi  r_{+}^2 \left(-3 M r_{+}+3 q^2+\alpha ^2
r_{+}^4\right)\right)}{6 M r_{+}-9 q^2+\alpha ^2 r_{+}^4}.
\end{eqnarray}
Figure \textbf{9} (right plots) shows the graphical analysis of
corrected heat capacity versus $r_{+}$. For different values of
$\gamma$, the first plot shows that corrected heat capacity diverges
at $r_{+}=1.47$ which is the Davies point. The positive region
before the Davies point shows that BHs with small radii are stable
under the effect of thermal fluctuations while the negative region
after Davies point shows that BHs with large radii are unstable.
However, for increasing values of coupling parameter, critical
radius of RN BH with non-linear electrodynamics also increases (2nd
plot). We note that BHs with small radii are unstable while BHs with
large radii are stable for different values of $\alpha$. However,
for different values of charge, only the divergence point of
corrected heat capacity changes while before and after the Davies
points, it remains in equilibrium condition (3rd plot).
\begin{figure}\center
\epsfig{file=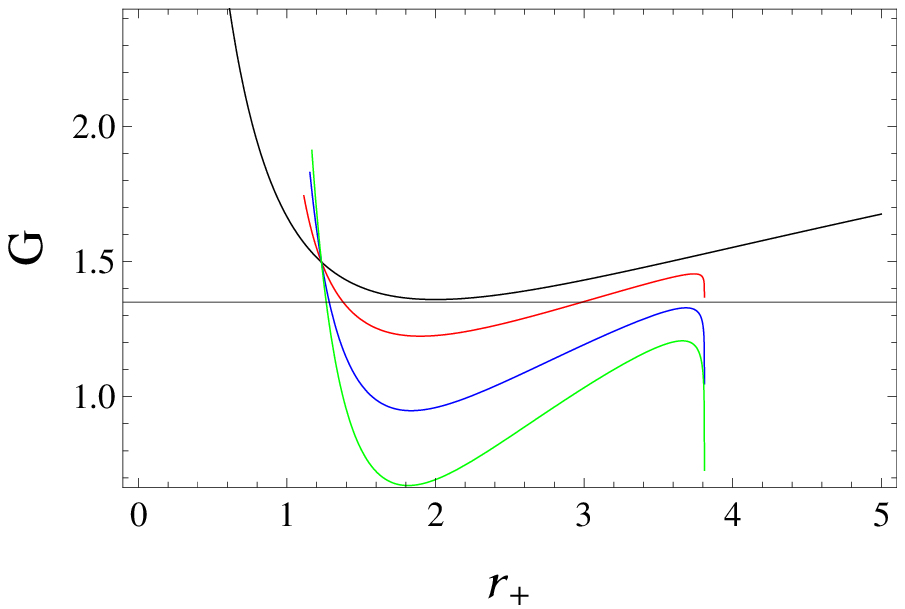,width=0.5\linewidth}\epsfig{file=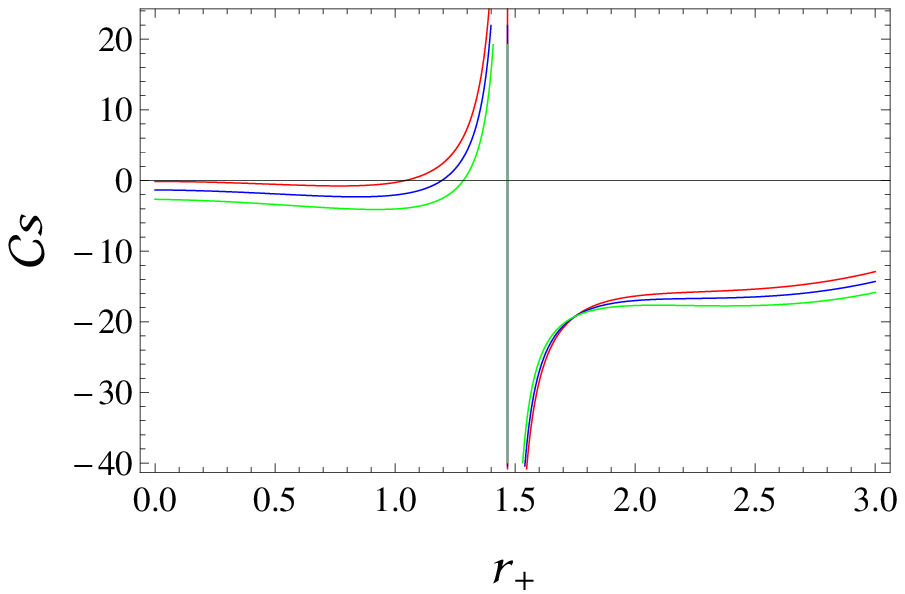,width=0.5\linewidth}
\epsfig{file=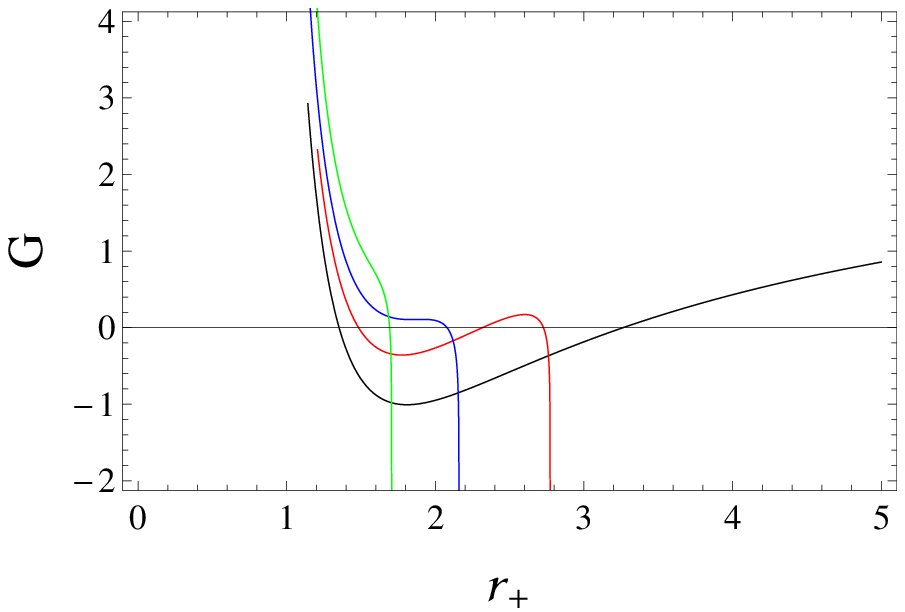,width=0.5\linewidth}\epsfig{file=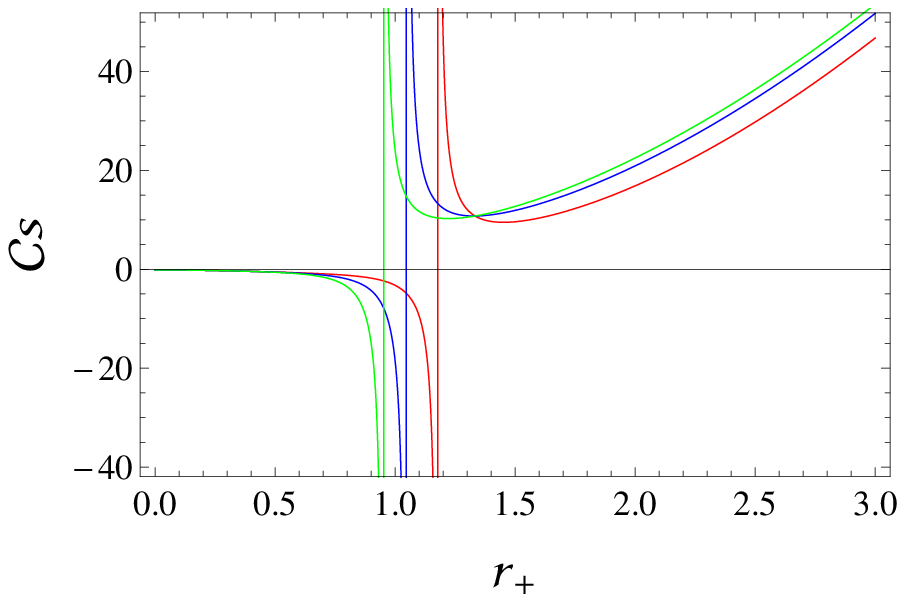,width=0.5\linewidth}
\epsfig{file=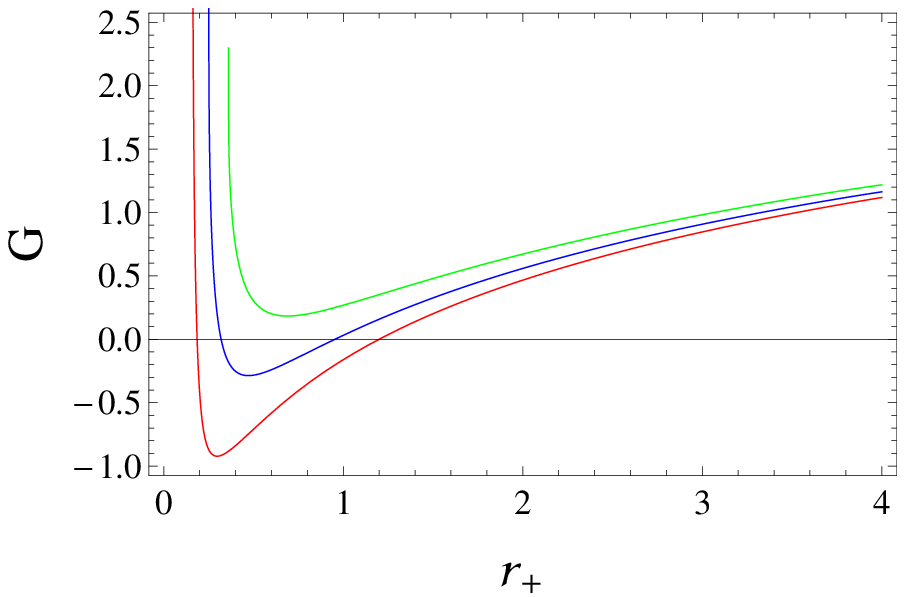,width=0.5\linewidth}\epsfig{file=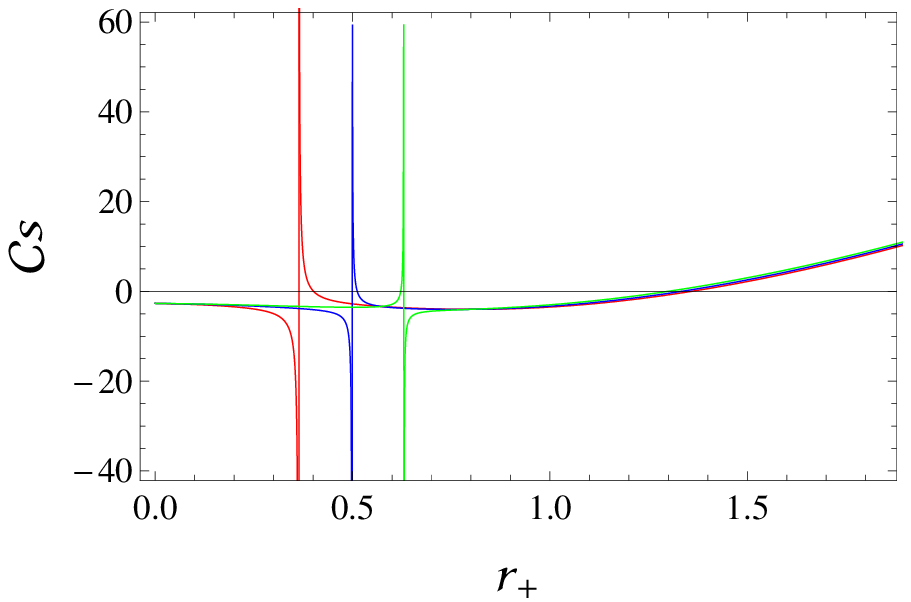,width=0.5\linewidth}
\caption{Gibbs free energy (left 3 plots) versus $r_{+}$ with
$M=q=1$. We take $\alpha=0.2$ with $\gamma=0$ (black), 1(red),
3(blue), 5(green) for the 1st plot, $\gamma=15$ with
$\alpha=0$(black), 0.3(red), 0.4(blue), 0.4(green) for the 2nd plot
and $M=1$, $\alpha=\gamma=0.2$ with $q=0.4$(red), 0.5(blue),
0.6(green) for the 3rd plot. Corrected heat capacity (right 3 plots)
versus $r_{+}$ for $M=q=1$. We take $\alpha=0.2$ with
$\gamma=0.1$(red), 1(blue), 2(green) for the 1st plot, $\gamma=0.1$
with $\alpha=1$(red), 1.5(blue), 2(green) for the 2nd plot and
$M=1$, $\alpha=\gamma=2$ with $q=0.5$(red), 0.6(blue), 0.7(green)
for the 3rd plot.}
\end{figure}

We can also analyze the stability of a BH by using trace of Hessian
matrix. The Hessian matrix contains the partial derivatives of
Helmholtz free energy with respect to Hawking temperature as well as
chemical potential $\upsilon=\Big(\frac{\partial M}{\partial
q}\Big)_{r_{+}}$. The Hessian matrix is given as \cite{14}
\begin{equation*}\label{1}
H=\left(
\begin{array}{cc}
H_{ii} & H_{ij} \\
H_{ji} & H_{jj} \\
\end{array}
\right) = \left( \begin{array}{cc} \frac{\partial^{2} F}{\partial
T^{2}} & \frac
{\partial^{2} F}{\partial T \partial \upsilon} \\
\frac{\partial^{2} F}{\partial \upsilon \partial T } &
\frac{\partial^{2} F}{\partial \upsilon^{2}} \\
\end{array}
\right),
\end{equation*}
with $i \neq j$, and
\begin{equation}\nonumber
H_{ii}=\frac{\partial^{2} F}{\partial T^{2}},\quad
H_{ij}=\frac{\partial^{2} F}{\partial T \partial \upsilon},\quad
H_{ji}=\frac{\partial^{2} F}{\partial \upsilon \partial T },\quad
H_{jj}=\frac{\partial^{2} F}{\partial \upsilon^{2}}.
\end{equation}
It is noted that $H_{ii}H_{jj}=H_{ij}H_{ji}$, which impies that one
of the eigenvalues of Hessian matrix is zero. Therefore, to
determine the stability of the considered geometry, we use trace of
the Hessian matrix as
\begin{equation}\label{34}
Tr(H)=H_{ii}+H_{jj},
\end{equation}
where
\begin{eqnarray}\nonumber
H_{ii}&=&-\left(6 \pi  r_{+}^4 \left(\frac{2 \gamma  \left(3 M
r_{+}-6 q^2+2 \alpha ^2 r_{+}^4\right)}{r_{+} \left(-3 M r_{+}+3
q^2+\alpha ^2 r_{+}^4\right)}-2 \pi
r_{+}\right)\right)\\\nonumber&\times&\left(6 M r_{+}-9 q^2+\alpha
^2 r_{+}^4\right)^{-1},\\\nonumber H_{jj}&=&\left(\frac{2 \gamma
\left(q+\alpha  r_{+}^2\right) \left(-3 M r_{+}+6 q^2-2 \alpha ^2
r_{+}^4\right) \left(-6 M r_{+}+9 q^2-\alpha ^2 r_{+}^4\right)}{3 M
r_{+}-3 q^2-\alpha ^2 r_{+}^4}\right.\\\nonumber&+&\left.\pi \left(3
\alpha r_{+}^4 \left(9 q^2-4 M r_{+}\right)+9 q^3 r_{+}^2+3 \alpha
^2 q r_{+}^6+\alpha ^3 r_{+}^8\right)+\gamma \left(-12 M q
r_{+}\right.\right.\\\nonumber&-&\left.\left.24 \alpha  M r_{+}^3+27
q^3+45 \alpha q^2 r_{+}^2+\alpha ^2 q r_{+}^4-\alpha ^3
r_{+}^6\right) \left(\ln \left(36 \pi
r_{+}^4\right)\right.\right.\\\nonumber&-&\left.\left.2 \ln \left(3
M r_{+}-3 q^2-\alpha ^2 r_{+}^4\right)\right)\right)\left(6 \pi
r_{+}^3 \left(q+\alpha r_{+}^2\right)^3\right)^{-1},
\end{eqnarray}
and we obtain
\begin{eqnarray}\nonumber
Tr&=&\left(\frac{2 \gamma  \left(q+\alpha  r_{+}^2\right) \left(-3 M
r_{+}+6 q^2-2 \alpha ^2 r_{+}^4\right) \left(-6 M r_{+}+9 q^2-\alpha
^2 r_{+}^4\right)}{3 M r_{+}-3 q^2-\alpha ^2
r_{+}^4}\right.\\\nonumber&+&\left.\pi \left(3 \alpha r_{+}^4
\left(9 q^2-4 M r_{+}\right)+9 q^3 r_{+}^2+3 \alpha ^2 q
r_{+}^6+\alpha ^3 r_{+}^8\right)+\gamma \left(-12 M q
r_{+}\right.\right.\\\nonumber&-&\left.\left.24 \alpha  M r_{+}^3+27
q^3+45 \alpha q^2 r_{+}^2+\alpha ^2 q r_{+}^4-\alpha ^3
r_{+}^6\right) \left(\ln \left(36 \pi
r_{+}^4\right)\right.\right.\\\nonumber&-&\left.\left.2 \ln \left(3
M r_{+}-3 q^2-\alpha ^2 r_{+}^4\right)\right)\right)\left(6 \pi
r_{+}^3 \left(q+\alpha r_{+}^2\right)^3\right)^{-1}-\left(6 \pi
r_{+}^4\right.\\\nonumber&\times&\left. \left(\frac{2 \gamma \left(3
M r_{+}-6 q^2+2 \alpha ^2 r_{+}^4\right)}{r_{+} \left(-3 M r_{+}+3
q^2+\alpha ^2 r_{+}^4\right)}-2 \pi r_{+}\right)\right)\left(6 M
r_{+}-9 q^2+\alpha ^2 r_{+}^4\right)^{-1}.
\end{eqnarray}
A thermodynamical system is said to be stable if $Tr(H)\geq0$
\cite{39}. Figure \textbf{10} shows the graphical analysis of
Hessian trace in terms of $r_{+}$ for different values of correction
parameter, coupling parameter and charge. It is observed that BHs
with small radii are unstable while BHs of large radii are stable
for considered parameters which is same as noticed in the graphical
analysis of corrected heat capacity (Figure \textbf{9}).
\begin{figure}\center
\epsfig{file=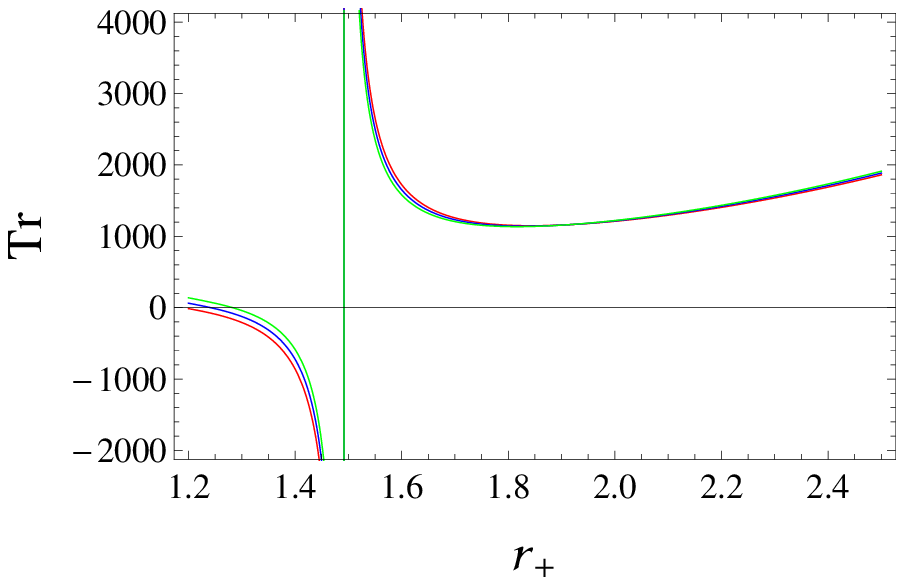,width=0.5\linewidth}\epsfig{file=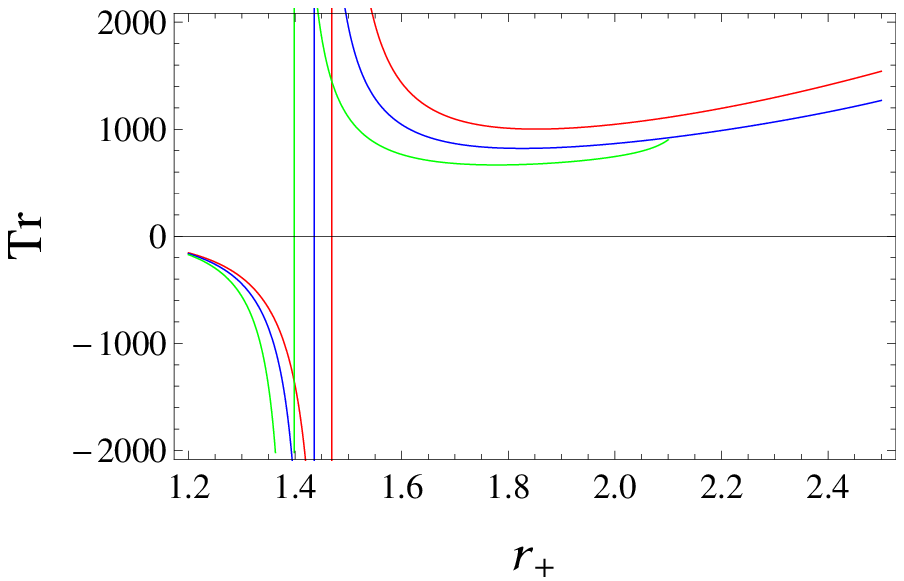,width=0.5\linewidth}
\epsfig{file=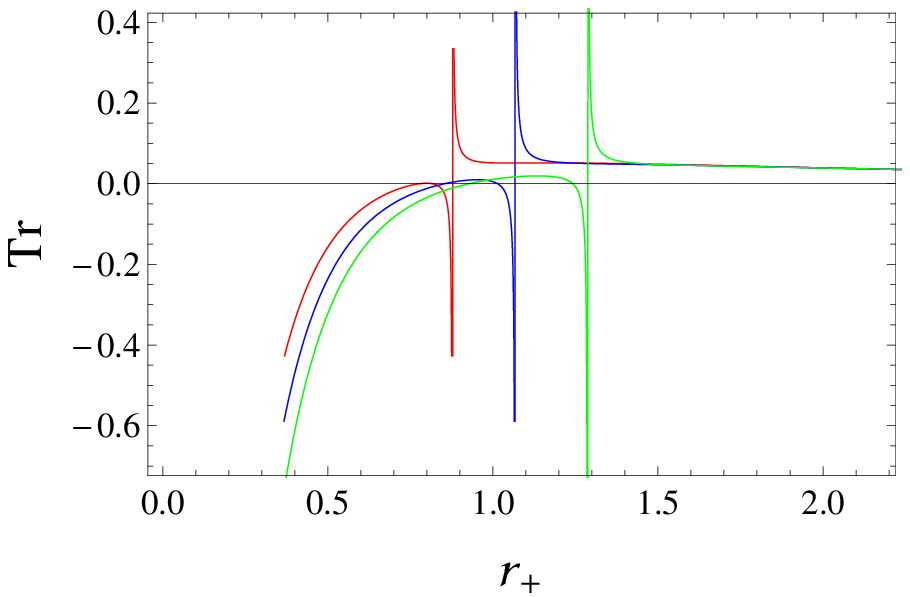,width=0.5\linewidth}\caption{Hessian trace
versus $r_{+}$ for $M=q=1$. We take $\alpha=0.1$ with
$\gamma=1$(red), 1.5(blue), 2(green) for the left plot, $\gamma=0.1$
with $\alpha=0.2$(red), 0.3(blue) and 0.4(green) for the right plot
and $M=1$, $\alpha=\gamma=0.2$ with $q=0.9$(red), 1(blue),
1.1(green) for the lower plot}.
\end{figure}

\section{Phase Transitions}

Here we examine the phase transitions of Hawking temperature as well
as heat capacity of RN BH with non-linear electrodynamic effects in
terms of entropy. The Hawking temperature evaluated in Eq.(\ref{3a})
can be written in the form of entropy as
\begin{equation}\nonumber
T=-\frac{\left(q-\alpha  r_{+}^2+r_{+}\right) (q-r_{+} (\alpha
r_{+}+1))}{4 r_{+} S}.
\end{equation}
Figure \textbf{11} (left plots) represents the phase transitions of
Hawking temperature in terms of $S$ for different values of $q$,
$r_{+}$ and $\alpha$, respectively. It is noted that the Hawking
temperature changes its phase from positive to negative for
increasing values of $q$ and $r_{+}$, respectively (1st and 2nd
plot). However, for increasing values of non-linear electrodynamic
parameter, the Hawking temperature changes its phase from negative
to positive (3rd plot). The expression of heat capacity for RN BH
with non-linear electrodynamic effects in terms of entropy can be
obtained from Eq.(\ref{4a}) as
\begin{equation}\nonumber
C=\frac{S \left(q-\alpha  r_{+}^2+r_{+}\right) (q-r_{+} (\alpha
r_{+}+1))}{-2 q^2+2 \alpha  q r_{+}^2+r_{+}^2}.
\end{equation}
The phase transitions of heat capacity versus entropy is shown in
Figure \textbf{11} (right plots) for different values of $q$,
$r_{+}$ and $\alpha$, respectively. It is noted that the right plots
of Figure \textbf{11} show that heat capacity of the considered BH
changes its phase from negative to positive for the increasing
values of $q$, $r_{+}$ and $\alpha$. It is also observed that
considered system is unstable for small values of considered
parameters while it shows stable behavior for large values of $q$,
$r_{+}$ and $\alpha$.
\begin{figure}\center
\epsfig{file=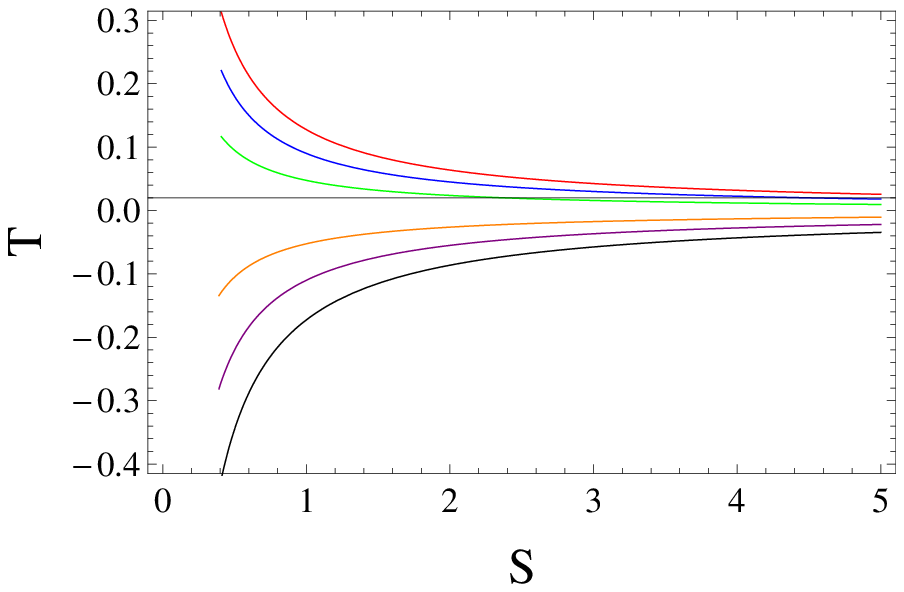,width=0.5\linewidth}\epsfig{file=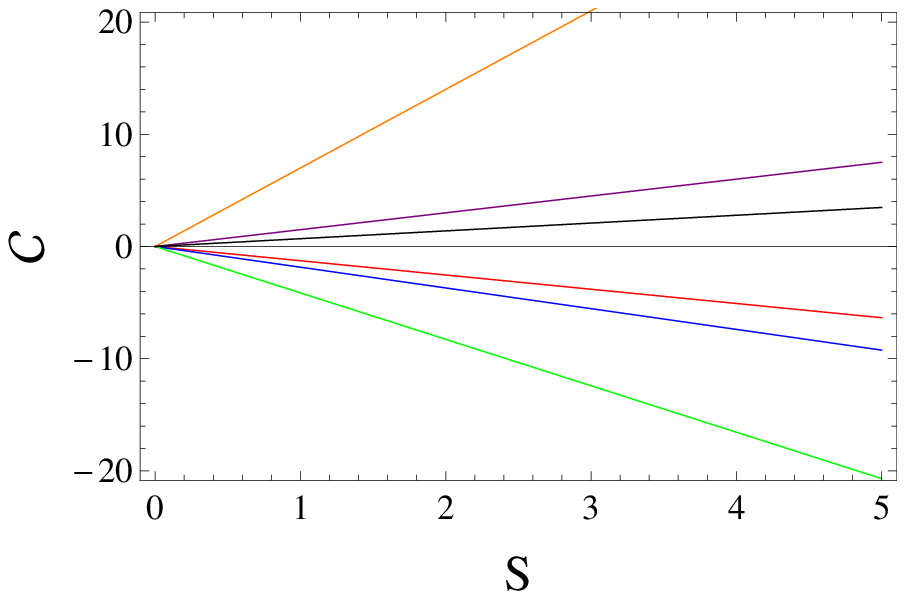,width=0.5\linewidth}
\epsfig{file=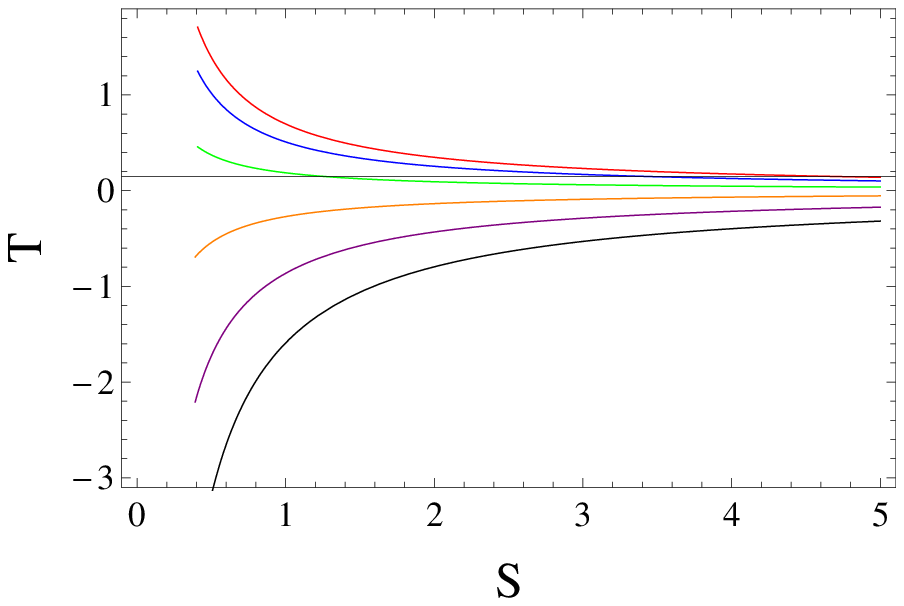,width=0.5\linewidth}\epsfig{file=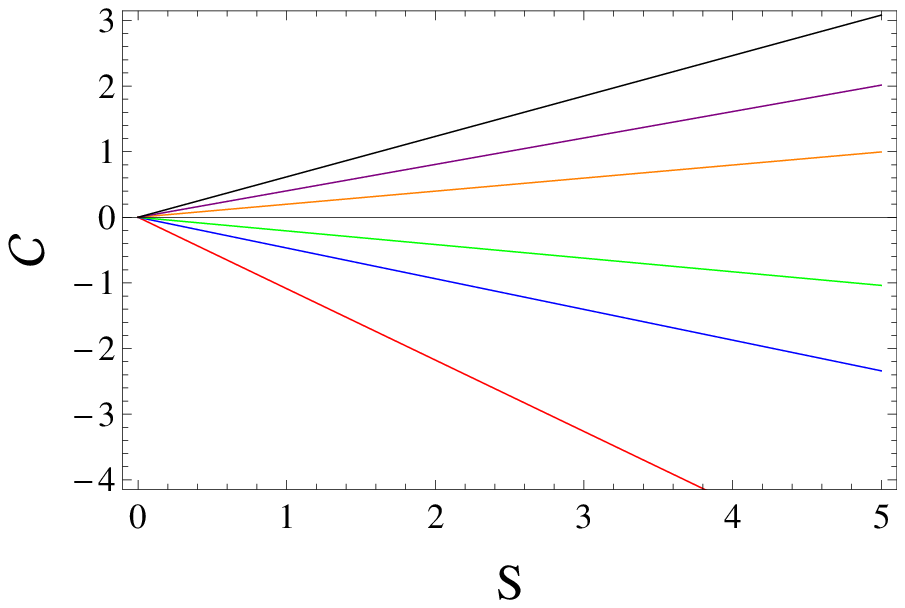,width=0.5\linewidth}
\epsfig{file=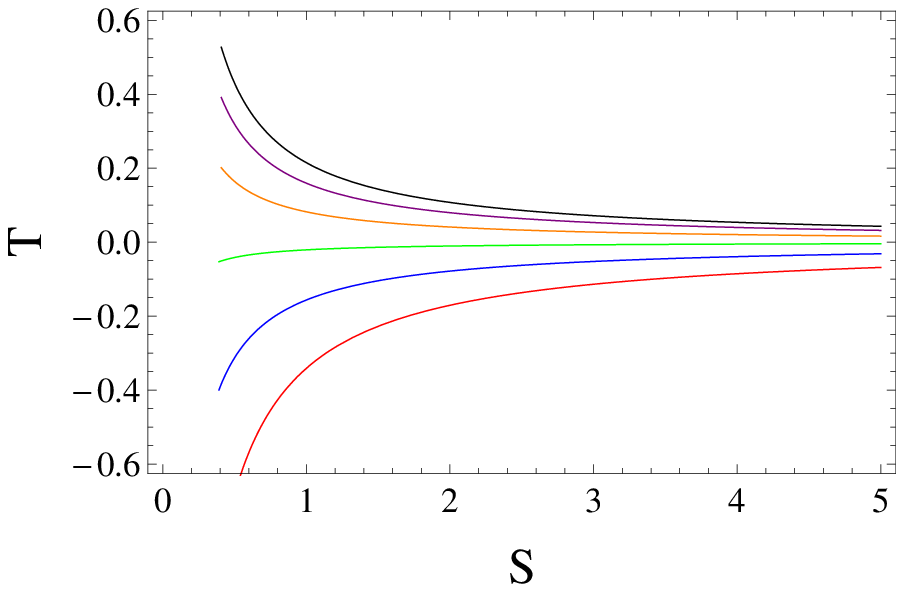,width=0.5\linewidth}\epsfig{file=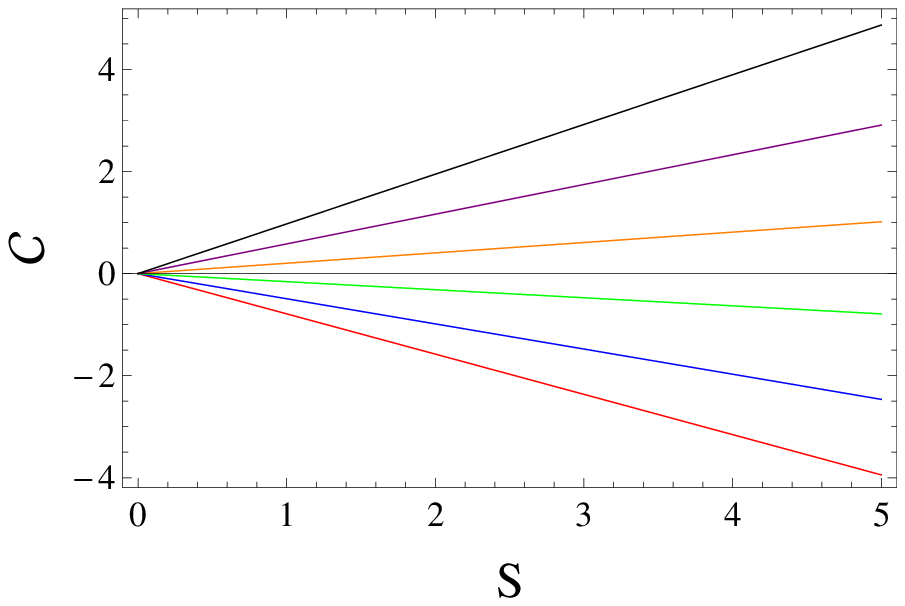,width=0.5\linewidth}
\caption{Hawking temperature (left 3 plots) versus $S$ for
$\alpha=1$, $r_{+}=1$. We take $q=1.7$(red), 1.8(blue), 1.9(green),
2.1(orange), 2.2(purple), 2.3(black) for the 1st plot, $\alpha=1$,
$q=1$ with $r_{+}=0.4$(red), 0.5(blue), 0.6(green), 0.7(orange),
0.8(purple), 0.9(black) for the 2nd plot and $r_{+}=3$, $q=1$ with
$\alpha=0.2$(red), 0.3(blue), 0.4(green), 0.5(orange), 0.6(purple),
0.7(black) for the 3rd plot. Heat capacity (right 3 plots) versus
$S$ with $r_{+}=1$, $\alpha=1$. We take $q=1.1$(red), 1.2(blue),
1.3(green), 1.4(orange), 1.5(purple), 1.6(black) for the 1st plot,
$\alpha=2$, $q=1$ with $r_{+}=0.7$(red), 0.8(blue), 0.9(green),
1.1(orange), 1.2(purple), 1.3(black) for the 2nd plot and $r_{+}=3$,
$q=1$, $\alpha=0.2$(red), 0.3(blue), 0.4(green), 0.5(orange),
0.6(purple), 0.7(black) for the 3rd plot.}
\end{figure}

\section{Concluding Remarks}

In this paper, we have discussed thermodynamical quantities for RN
BH with non-linear electrodynamic effects. We have then derived the
relationship between Davies point and QNMs as well as examined the
effects of thermal fluctuations on uncorrected thermodynamical
quantities and then compared both corrected as well as uncorrected
quantities graphically. We have also analyzed the phase transitions
of heat capacity and Hawking temperature.

Firstly, we have computed the Hawking temperature by means of
surface gravity with the relation $(T=\frac{\kappa}{2\pi})$ and then
heat capacity for the considered BH. The graphical analysis of heat
capacity shows that heat capacity diverges at $r_{+}=1.2$ (Figure
\textbf{2}). It is also found that BHs with small radii are stable
in the presence of non-linear electrodynamic effects while BHs with
large radii are unstable. Secondly, we have discussed the null
geodesics and QNMs for which we have evaluated the real part as
angular velocity and imaginary part as Lyapunov exponent of QNMs. It
is observed that angular velocity of photon remains unaffected by
the effect of coupling parameter while lyapunov exponent shows
increases behavior and significantly affected by coupling parameter.
The interesting fact is that the Davies point evaluated from plots
of heat capacity (Figure \textbf{3}) and angular velocity (Figure
\textbf{4}) are exactly the same.

Further, we have obtained the corrected expression for entropy by
considering logarithmic corrections of first order and then used it
to calculate the modified results of thermodynamical potentials. It
is found that corrected entropy remains positive as well as
increasing monotonically but remains unaffected for increasing
values of correction parameter. However, for increasing values of
non-linear electrodynamics parameter, the corrected entropy of small
as well as large BHs have similar behavior. The corrected Hawking
temperature changes its behavior from negative to positive for
increasing values of correction parameter and coincides with the
equilibrium condition for large values of $r_{+}$. However, it
becomes positive valued function as well as increases gradually for
higher values of $\alpha$ and large BHs are affected more by
$\alpha$ as compared to small ones.

The Helmholtz free energy shows similar behavior in the presence as
well as absence of correction parameter but logarithmic corrections
affect small BHs more as compared to large ones. However, for
increasing values of $\alpha$, Helmholtz free energy shows
decreasing behavior but it affects BHs with large radii more as
compared to BHs of small radii. The internal energy fluctuates in
both absence as well as presence of $\gamma$ and affects both large
as well as small BHs negligibly. The internal energy of BHs with
radii from $r_{+}=0$ to 4 remains unaffected for increasing values
of $\alpha$ and only affects BHs with radii $r_{+}>4$.

The enthalpy of the considered system also fluctuates with
increasing values of $\gamma$ and  decreases continuously for large
event horizon. However, this shows decreasing behavior for small
values of $r_{+}$ and increases with large values of $r_{+}$ for
$\alpha=0$ but shows only decreasing trend in the presence of
$\alpha$. It is also noted that the increasing values of both
correction and coupling parameters only affect BHs with large radii.
We have found that the corrected heat capacity diverges at
$r_{+}=1.47$ and also observed that BHs with radii greater than 1.47
are unstable under the effect of thermal fluctuations. Black holes
with small radii are unstable while BHs with large radii are stable
for considered values of coupling parameter. It is also noted that
critical radius of BH increases with the increasing values of
$\alpha$.

Finally, we have investigated the phase transitions of Hawking
temperature as well as heat capacity in terms of entropy for the
considered BH. It is found that Hawking temperature changes its
phase from positive to negative for increasing values of both charge
and horizon radius. This shows that BHs with small charge and radii
are more hotter than the BHs with large charge and radii while it
shows opposite behavior for increasing values of coupling parameter.
The heat capacity changes its phase from negative to positive for
increasing values of considered parameters which shows that BHs with
small values of $q$, $r_{+}$ and $\alpha$ are unstable while BHs
with large values are stable throughout the considered domain. We
would like to mention here that all our results reduce to the RN BH
for $\alpha=0$ \cite{13}.

\end{document}